\documentclass[review]{elsarticle}
\usepackage{multirow} 
\usepackage{xcolor}
\usepackage{lineno,hyperref}
\modulolinenumbers[5]
\usepackage{subcaption}

\usepackage{blindtext}
\usepackage{booktabs}
\usepackage{amsbsy} 
\usepackage{upgreek}

\begin{document}
 
\begin{frontmatter}

\title{Overcoming the Convergence Difficulty of Cohesive Zone Models through a Newton-Raphson Modification Technique}

\author[mymainaddress]{Reza Sepasdar}

\author[mymainaddress]{Maryam Shakiba\corref{mycorrespondingauthor}}
\cortext[mycorrespondingauthor]{Corresponding author}
\ead{mshakiba@vt.edu}

\address[mymainaddress]{Department of Civil and Environmental Engineering, Virginia Tech, 750 Drillfield Dr., Blacksburg, VA 24061, USA}

\begin{abstract} 
This paper studies the convergence difficulty of cohesive zone models in static analysis. It is shown that an inappropriate starting point of iterations in the Newton-Raphson method is responsible for the convergence difficulty. A simple, innovative approach is then proposed to overcome the convergence issue. The technique is robust, simple to implement in a finite element framework, does not compromise the accuracy of analysis, and provides fast convergence. The paper explains the implementation algorithm in detail and presents three benchmark examples. It is concluded that the method is computationally efficient, has a general application, and outperforms the existing methods.
\end{abstract}

\begin{keyword}
Cohesive Zone\sep Convergence Difficulty \sep Finite Element\sep Numerical Instability\sep Interface Debonding
\end{keyword}

\end{frontmatter}


\section{Introduction}
Cohesive zone models (CZMs) are popular methods to simulate fracture in engineering applications. Delamination of layered composites, crack propagation in materials, and debonding of fiber/matrix interfaces in fiber-reinforced composites are examples of CZMs applications  \textcolor{blue}{\cite{roe2003irreversible,aymerich2008prediction,carloni2012experimental,herraez2015transverse,turon2018accurate,shakibatransverse}}. The cohesive models are easy to implement within a finite element (FE) framework and adaptable to other nonlinear constitutive models. CZMs were also proven to be efficient, robust, and more accurate compared to continuum damage and fracture mechanics theories \textcolor{blue}{\cite{chaboche2001interface,hamitouche2008interface,dimitri2014consistency,yu2016viscous}}. However, cohesive models can encounter convergence difficulty in a static analysis once failure initiates \textcolor{blue}{\cite{volokh2004comparison,jiang2010cohesive,needleman2014some,zhu2015determination,simonovski2015cohesive}}. The convergence issue reduces CZMs efficiency and limits their range of applications.
 
Researchers studied the root of the convergence difficulty in CZMs and categorized it into two main groups of physical and numerical instability \textcolor{blue}{\cite{chaboche2001interface,gao2004simple,hamitouche2008interface,gu2015inertiaa,yu2016viscous}}. Several studies explained that fracture initiation and propagation is intrinsically a dynamic phenomenon. Thus, physical instability due to the lack of kinetic energy and inertial forces in static analysis causes the convergence issue \textcolor{blue}{\cite{hilber1977improved,hilber1978collocation,gao2004simple,gu2015inertiaa,michel2018new}}. Other studies, however, conjectured that numerical instability due to either a solution jump after failure or the presence of multiple solutions is responsible for the convergence difficulty \textcolor{blue}{\cite{chaboche2001interface,yu2016viscous}}. Several methods were consequently developed to resolve the convergence difficulty based on the hypothesized root of the issue.

The most common existing methods to overcome the convergence difficulty in CZMs are viscous regularization \textcolor{blue}{\cite{chaboche2001interface,gao2004simple}}, decrease of global displacement (known as Riks method) \textcolor{blue}{\cite{riks1979incremental}}, and full dynamic analysis \textcolor{blue}{\cite{hilber1977improved,hilber1978collocation}}. The different points of view on the convergence difficulty of CZMs and their performances for static frameworks are explained in detail in Section \ref{subsec: Convergence Difficulty}. In summary, the available methods to overcome the convergence difficulty in static analysis make the computations expensive, compromise the accuracy, or require complex programming. Likewise, full dynamic analysis (implicit or explicit) is computationally very costly. Therefore, it is imperative to revisit and fully understand the root of the convergence difficulty and propose a method to resolve this issue efficiently.   

In this paper, the behavior of CZMs under static loading and their convergence issue in an FE analysis are comprehensively studied through one-dimensional (1-D) and two-dimensional (2-D) representative examples. The Newton-Raphson (NR) iteration at the moment of instability based on a new point of view is analyzed. The NR analysis sheds light on the actual root of the convergence difficulty. A novel and simple method for overcoming the convergence issue is then presented and proven to be efficient and robust. All concepts presented in this research, as well as the proposed method, can easily be generalized to three-dimensional (3-D) analyses.

This paper is organized as follows. The behavior of cohesive interfaces, their implementation within an FE framework, different points of view on the convergence difficulty, and the existing solutions in the literature are discussed in Section~\ref{sec: Cohesive Zone Model Behavior}. The actual root of the convergence difficulty is then determined based on the examination of the NR iterations at the moment of instability in Section~\ref{sec: Analysis of CZMs' Convergence Difficulty}. Subsequently, a novel method is proposed in Section~\ref{proposed approach} to overcome the convergence issue. In the end, three application examples are presented to demonstrate the validity, robustness, and efficiency of the proposed method.

\section{Cohesive Zone Model Behavior} \label{sec: Cohesive Zone Model Behavior}
\begin{figure*}
    \centering
	\begin{subfigure}{0.6\textwidth}
	    \centering
        \includegraphics[width=\textwidth]{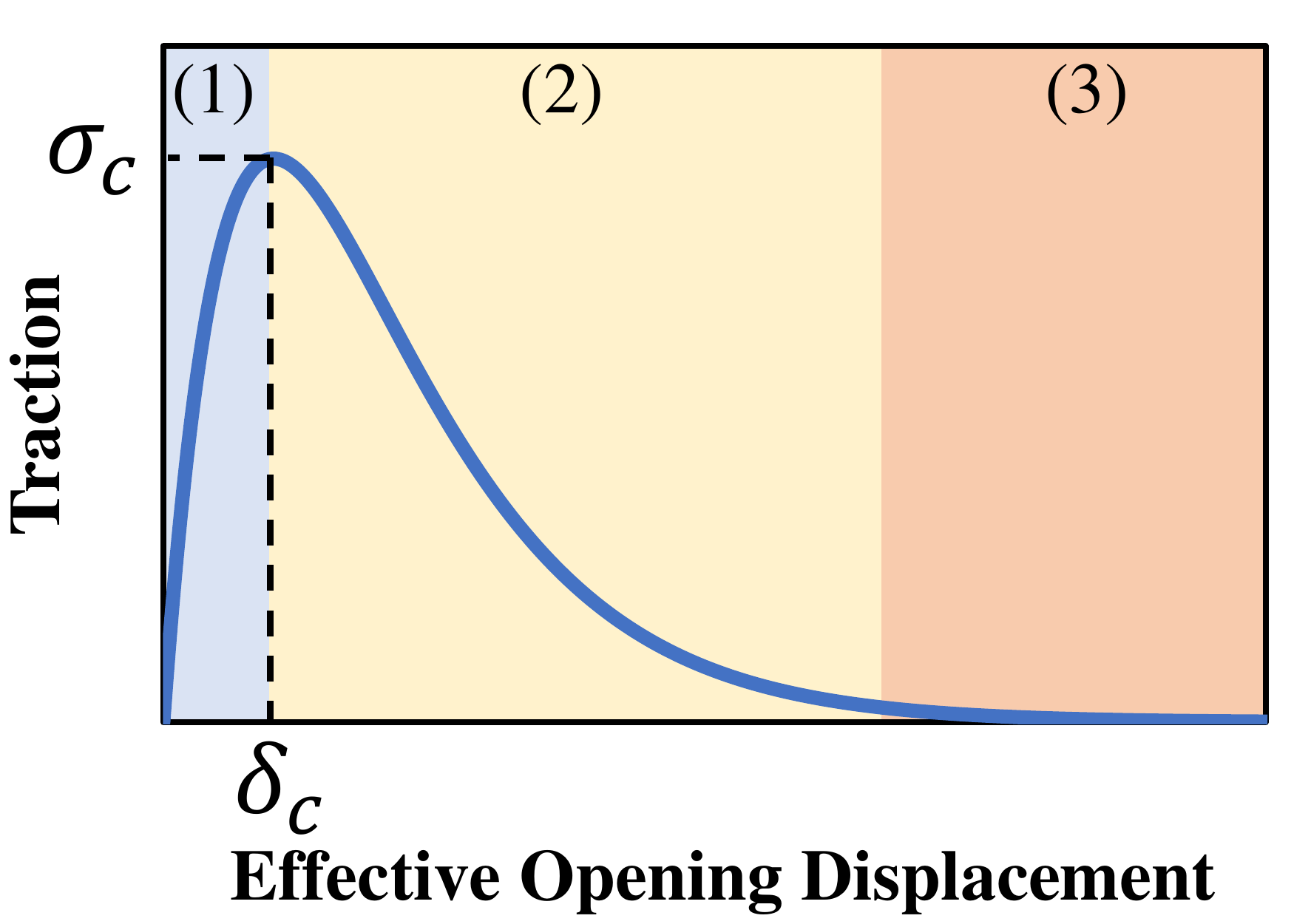}
        \caption{} \label{fig:1a}
    \end{subfigure}
	\begin{subfigure}{0.27\textwidth}
	    \centering
        \includegraphics[width=\textwidth]{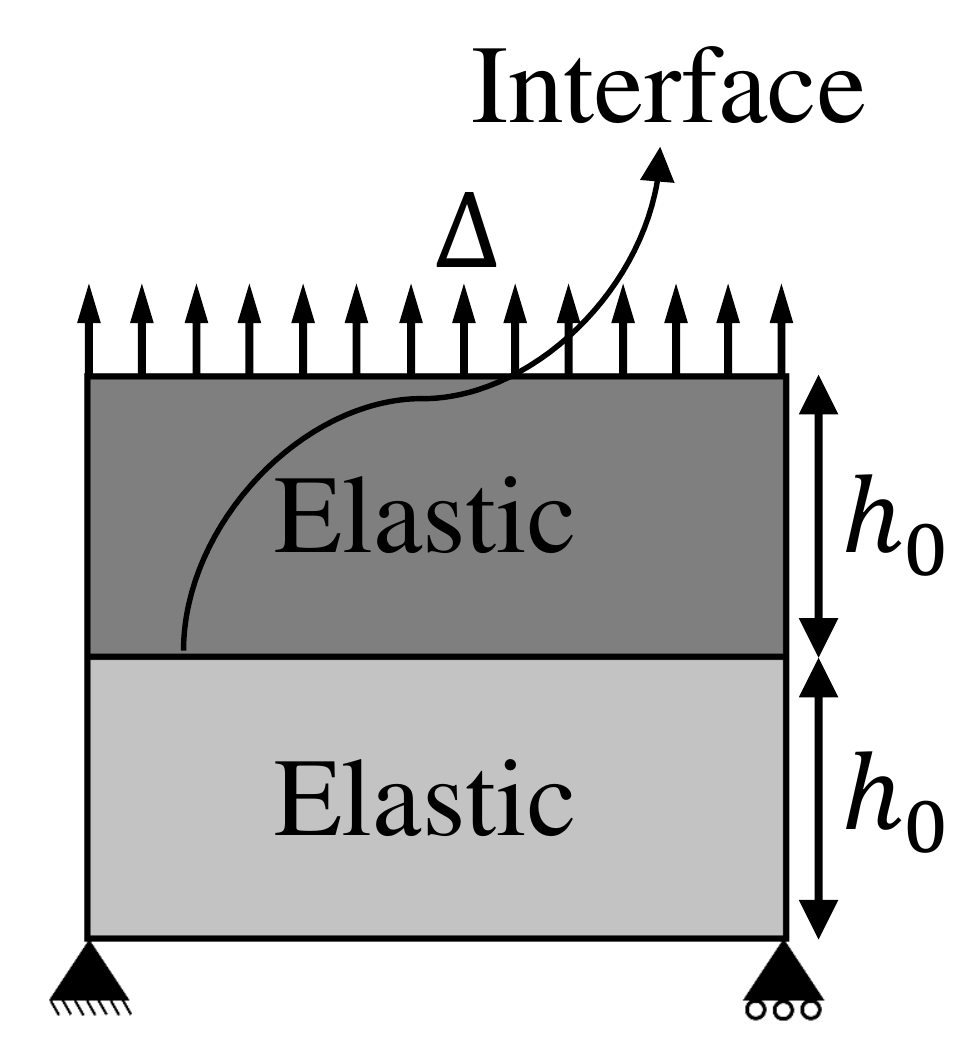} 
        \caption{} \label{fig:1b}
    \end{subfigure}
	\\
	\begin{subfigure}{0.33\textwidth}
	    \centering
        \includegraphics[width=\textwidth]{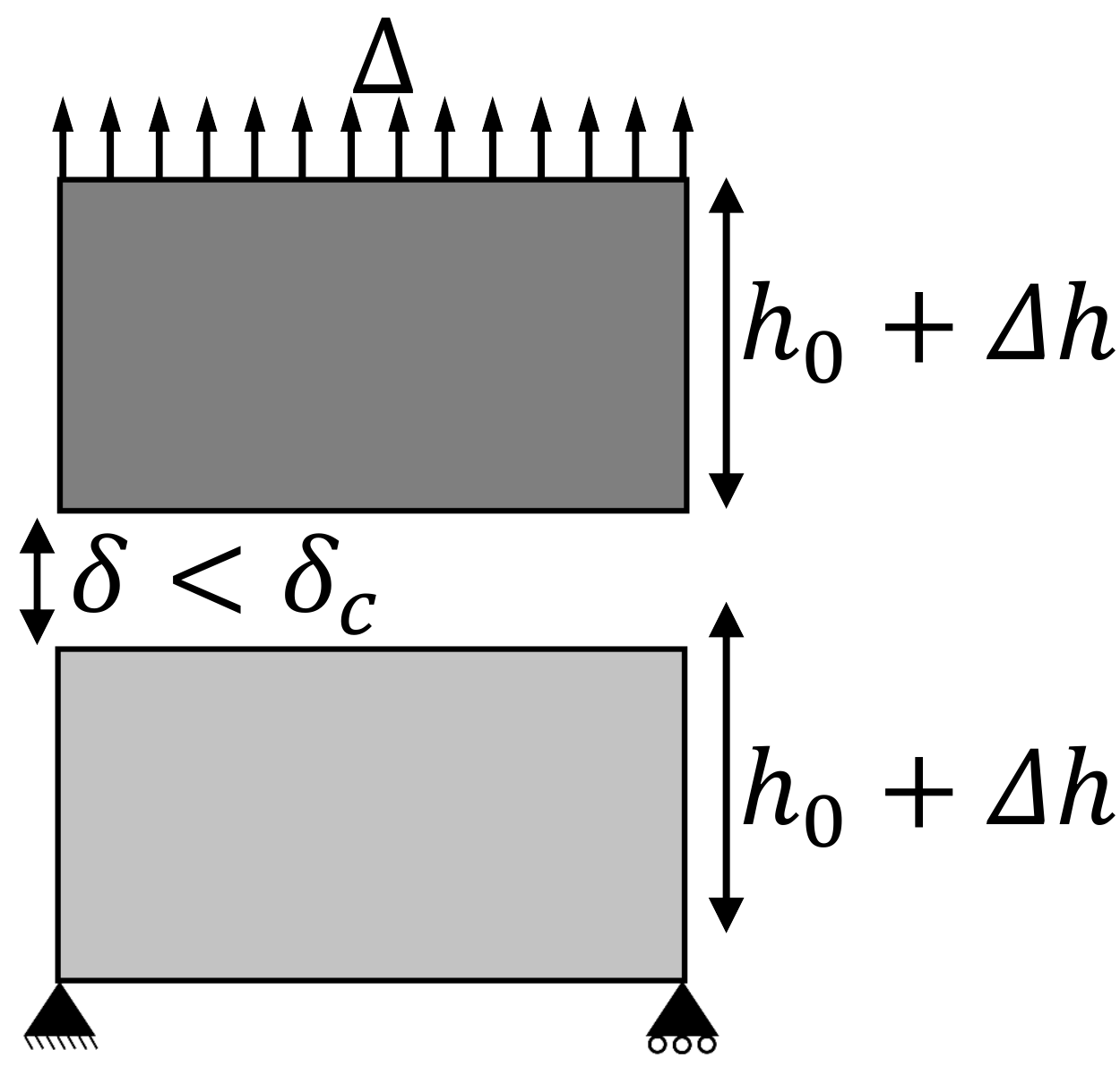}
        \caption{} \label{fig:1c}
    \end{subfigure}
	\begin{subfigure}{0.24\textwidth}
	    \centering
        \includegraphics[width=\textwidth]{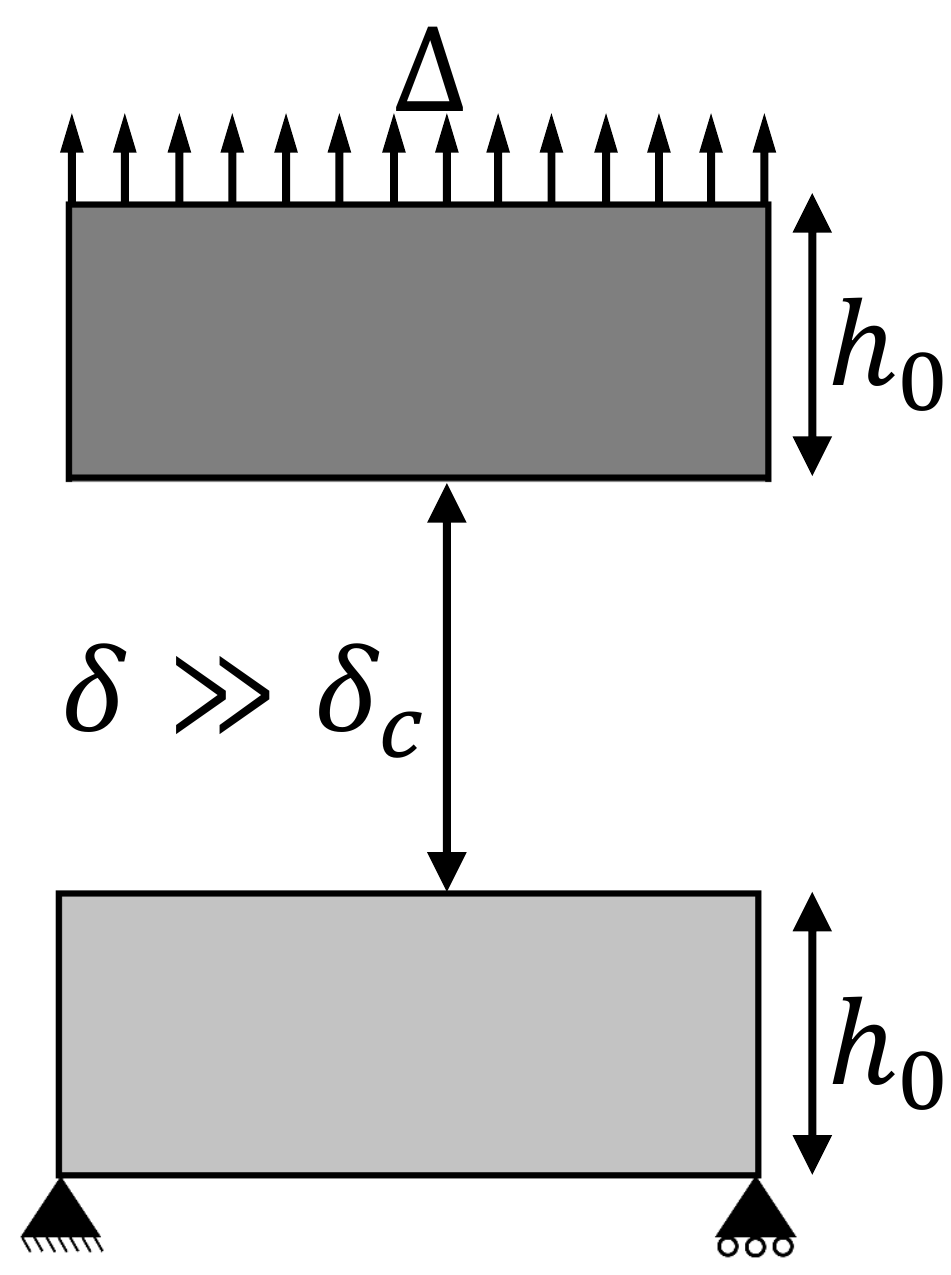}
        \caption{} \label{fig:1d}
    \end{subfigure}
    \begin{subfigure}{0.41\textwidth}
	    \centering
        \includegraphics[width=\textwidth]{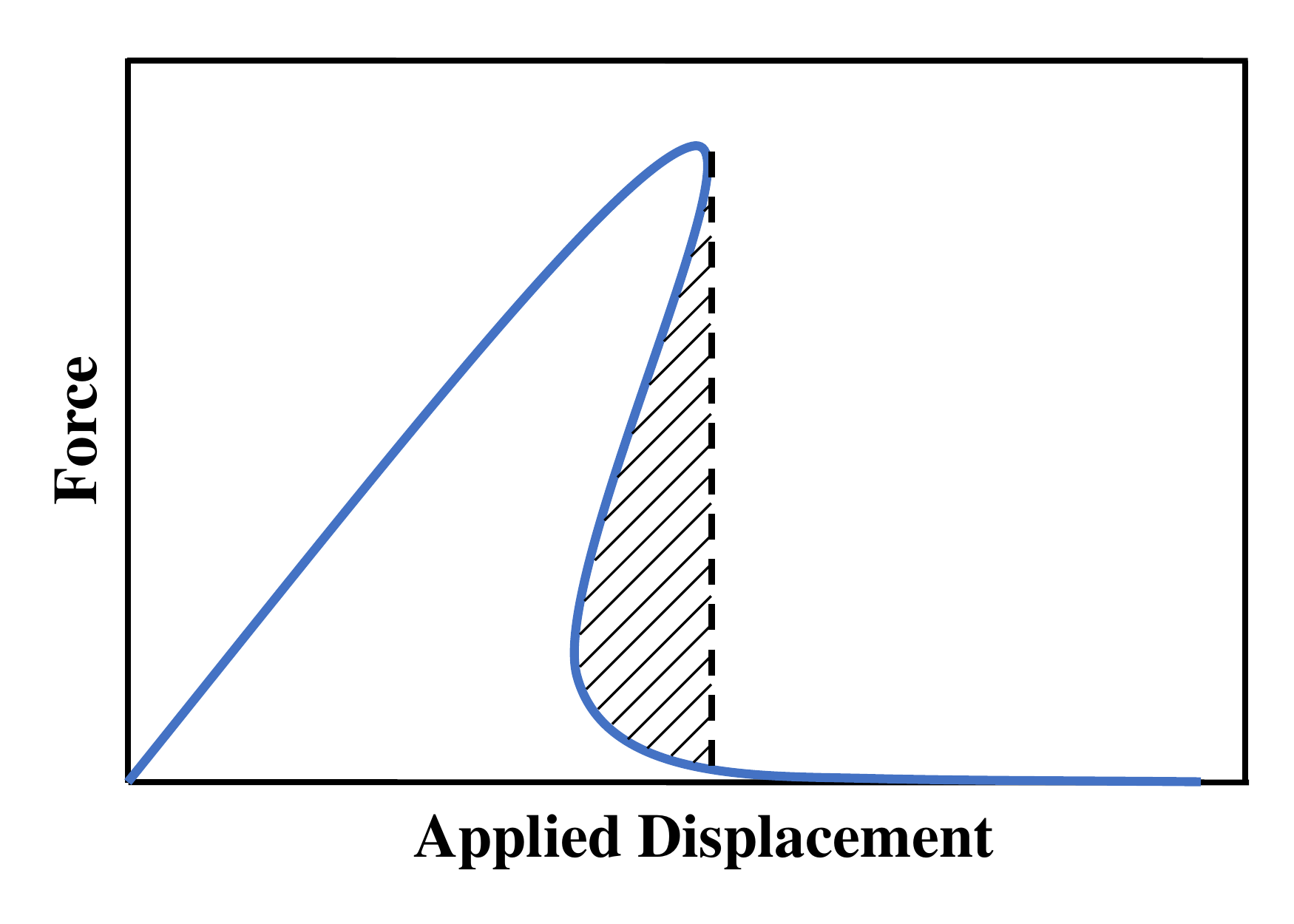}
        \caption{} \label{fig:1e}
    \end{subfigure}

	\caption{(a) A schematic representation of an exponential CZM and three stages of an interface debonding behavior, and (b), (c), (d) demonstrate debonding stages of a simple elastic bar with an interface, and (e) a schematic representation of a snap-back resulting from the decrease of applied displacement.}
	\label{FIG:exponential behavior and loading stages}
\end{figure*}

Several CZMs were developed in the literature to express the interface behavior based on application problems. The models use nonlinear functions, including exponential \textcolor{blue}{\cite{needleman1990analysis,ortiz1999finite,chandra2002some}}, bilinear \textcolor{blue}{\cite{hillerborg1976analysis,mi1998progressive,alfano2001finite}}, trapezoidal \textcolor{blue}{\cite{tvergaard1992relation,tvergaard1993influence}}, parabolic \textcolor{blue}{\cite{allix1995damage,allix1996modeling}}. Various functions may have different performances and convergence rates depending on the mesh size, geometry, and material properties in an FE simulation \textcolor{blue}{\cite{volokh2004comparison,alfano2006influence}}; however, their general behavior is similar. Thus, an exponential function as a typical traction-separation or debonding law is adopted in this work to study the behavior of CZMs. The exponential function was chosen as it merely provides a better visual presentation of the idea in this work. In the exponential model \textcolor{blue}{\cite{ortiz1999finite}}, the traction in the interface, $\sigma$, is related to the effective opening displacement, $\delta$, by
\begin{eqnarray}\label{eq:1}
\sigma=\frac{\sigma_c}{\delta_c}\delta\: e^{(1-\delta/\delta_c)}
\end{eqnarray}
where $\sigma_c$ is the cohesive strength of the interface corresponding to a critical opening displacement $\delta_c$. This exponential CZM is schematically presented in \textcolor{blue}{Fig.~\ref{FIG:exponential behavior and loading stages}~(a)}.

A few models are available in the literature to calculate CZMs' effective opening displacement in the case of mixed-mode loading, which account for the interaction between normal and tangential behavior of interfaces \textcolor{blue}{\cite{tvergaard1990effect,tvergaard1993influence,ortiz1999finite,geubelle1998impact}}. In this paper, the coupling between the normal and tangential debonding is defined based on a proposed relation by Ortiz and Pandolfi \textcolor{blue}{\cite{ortiz1999finite}}
\begin{eqnarray}\label{eq:2}
\delta=\sqrt{\delta_n^2+\beta\delta_t^2}
\end{eqnarray}
where $\beta$ is a factor between 0 and 1 that differentiates between the normal and tangential interface debonding. If $\beta=1$, $\delta$ becomes the true distance between the pair nodes of the interfaces. 

The CZM is presented in 1-D in this section but can be easily extended to 3-D cases. It must be noted that the concepts and methods presented in this paper have a general functionality and are independent of the types of CZMs' behavior.

\subsection{Convergence Difficulty} \label{subsec: Convergence Difficulty} 
To study the behavior of cohesive interfaces, a simple elastic bar with an interface in between, as illustrated in \textcolor{blue}{Fig.~\ref{FIG:exponential behavior and loading stages}~(b)}, is considered. The elastic bar is restrained at the bottom while a static monotonic displacement, $\Delta$, is applied at its top boundary. The applied displacement on this bar is equal to the induced opening displacement in the interface, $\delta$, in addition to the deformation in the elastic bar, $2\Delta h$. By applying the displacement on the elastic bar, the interface starts to debond while its traction increases. \textcolor{blue}{Fig.~\ref{FIG:exponential behavior and loading stages}~(c)} schematically illustrates this debonding initiation stage, which corresponds to phase one in \textcolor{blue}{Fig.~\ref{FIG:exponential behavior and loading stages}~(a)}.
Once the interface fails (i.e., $\delta > \delta_c$), the interface traction decreases while its opening displacement increases. In this stage, elastic deformation in the bar recovers, which is related to phase two in \textcolor{blue}{Fig.~\ref{FIG:exponential behavior and loading stages}~(a)}. By increasing the applied displacement, the interface traction decreases until it converges to zero, which corresponds to phase three in \textcolor{blue}{Fig.~\ref{FIG:exponential behavior and loading stages}~(a)}. The elastic parts, then, completely recover to their initial undeformed shape. Any further increase in the applied displacement is followed by a rigid body movement of the top elastic part, as it can be seen in \textcolor{blue}{Fig.~\ref{FIG:exponential behavior and loading stages}~(d)}. 

It is known that if the elastic material's stiffness is low compared to the interface's local tangent stiffness, the failure happens instantaneously and causes instability in the numerical solution \textcolor{blue}{\cite{chaboche2001interface}}. An instantaneous failure of the interface implies that the interface traction quickly drops to zero, which is followed by a sudden deformation recovery in the elastic parts. This phenomenon is known as a solution jump, which causes convergence difficulty in static analyses \textcolor{blue}{\cite{yu2016viscous,chaboche2001interface}}. Several researchers explained the sudden recovery as a dynamic phenomenon and represented it by a snap-back in the total force-displacement response as it is schematically shown in  \textcolor{blue}{Fig.~\ref{FIG:exponential behavior and loading stages}~(e)} \textcolor{blue}{\cite{yu2016viscous,gao2004simple,michel2018new,gu2015inertiaa}}. The snap-back portion results from the decrease in the applied displacement, and its area is argued to be equal to the missing kinetic energy in static analyses. After the snap-back portion is passed, the applied displacement can be increased again, and the convergence is easy afterward. Based on different reasoning on the roots of the convergence difficulty, several methods were proposed to resolve this issue. The most common methods are explained in the next subsection. 

\subsection{Existing Methods for Overcoming the Convergence Difficulty} \label{subsec: existing methods}
One possible solution to address the convergence difficulty in CZMs is to decrease the applied displacement. This solution typically uses the arc-length (Riks) iterative method due to its path-finding feature \textcolor{blue}{\cite{riks1979incremental,wang}}. The arc-length method requires additional displacement steps to capture the snap-back portion, which increases the computational cost. Besides, programming of such an algorithm is complex, especially when there are multiple local instabilities in the FE analysis. Moreover, Hamitouche et al., \textcolor{blue}{\cite{hamitouche2008interface}} and Yu et al., \textcolor{blue}{\cite{yu2016viscous}} argued that decreasing the applied displacement is not consistent with the nature of a displacement control analysis/experiment where the applied displacement always increases. 

Another common method to resolve CZMs' convergence difficulty is the well-known viscous regularization method  \textcolor{blue}{\cite{chaboche2001interface,gao2004simple}}. The viscous regularization method adds a so-called artificial viscosity term to the cohesive zone constitutive behavior to facilitate the convergence. The proponents of the physical instability as being the root of convergence difficulty argue that the artificial viscosity compensates for the missing kinetic energy in static analyses. 
Chaboche et al., \textcolor{blue}{\cite{chaboche2001interface}} and Yu et al., \textcolor{blue}{\cite{yu2016viscous}}, however, believed that the artificial viscosity would fix a numerical instability, which prevents the convergence of NR iterations.
Viscous regularization approach has gained more popularity compared to other methods due to the ease of implementation and its high convergence rate. The method, however, requires altering the cohesive constitutive behavior, which necessitates a sensitivity analysis on the effect of the artificial viscosity coefficient. It also requires additional displacement steps to simulate the instantaneous failure of the interface.

There are a few other proposed methods for addressing CZMs' convergence issue in the literature that are less popular due to their inefficiency and complexity. These methods are briefly reviewed here. One proposed approach is to reduce the mesh size as it decreases the convergence difficulty due to local instabilities  \textcolor{blue}{\cite{tvergaard1991micromechanical,chaboche2001interface}}. However, excessive mesh refinement decreases the efficiency of numerical analysis. Another alternative is to use quasi-Newton iterative methods, which do not have a quadratic convergence and are thus slow. Additional treatments were proposed to accelerate the convergence speed of quasi-Newton iterative methods \textcolor{blue}{\cite{michel2018new}}. Recently, Gu et al., \textcolor{blue}{\cite{gu2015inertiaa}} proposed the inclusion of inertial forces in the CZMs' constitutive behavior instead of artificial viscosity to compensate for the lack of kinetic energy. This method requires special programming, which involves solving a system of nonlinear equations for the increments with the convergence issue. There are also a few other methods in the literature to resolve the CZMs' convergence issue based on continuum damage mechanics \textcolor{blue}{\cite{peerlings1996gradient,de2006computational}}. 

In summary, the existent methods in the literature have limited applications due to their complexity, implementation difficulties, and increase in the computational costs \textcolor{blue}{\cite{chaboche2001interface}}. Therefore, it is necessary to comprehensively analyze the convergence difficulty related to CZMs and propose an efficient method to resolve this issue. 

\section{Analysis of CZMs' Convergence Difficulty} \label{sec: Analysis of CZMs' Convergence Difficulty}
In this section, the convergence difficulty of CZMs in static frameworks is studied through monitoring NR iterations at the moment of instability. It must be pointed out that the CZM, simple representative FE model, and discussions presented in this section are in a 2-D framework, but can easily be extended to 3-D analyses.

\begin{figure}
	\centering
	\begin{subfigure}{0.22\textwidth}
        	\includegraphics[width=\textwidth]{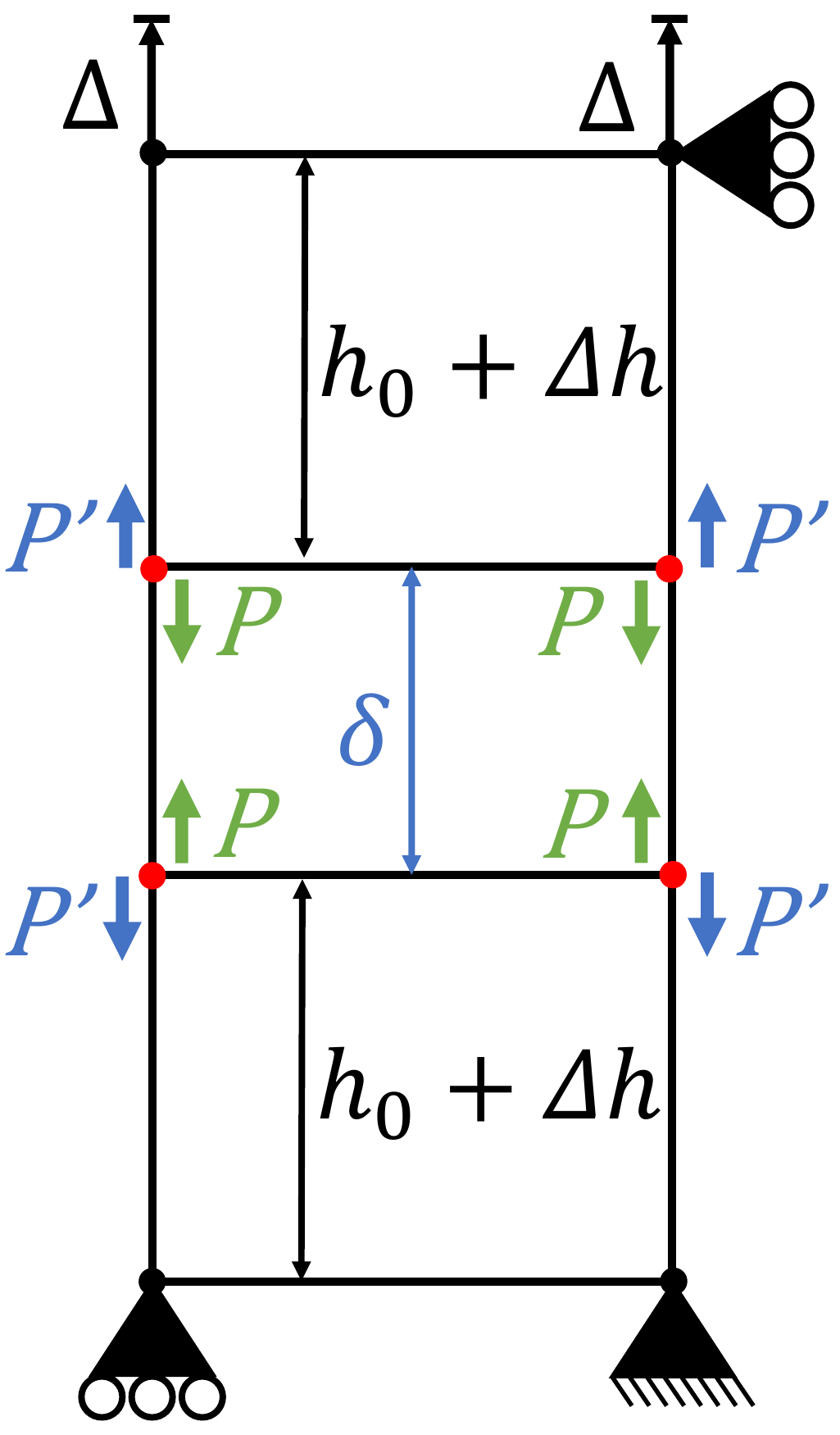}
        \caption{} \label{fig:2a}
    \end{subfigure}
    \\
    \begin{subfigure}{0.65\textwidth}
	    \centering
        \includegraphics[width=\textwidth]{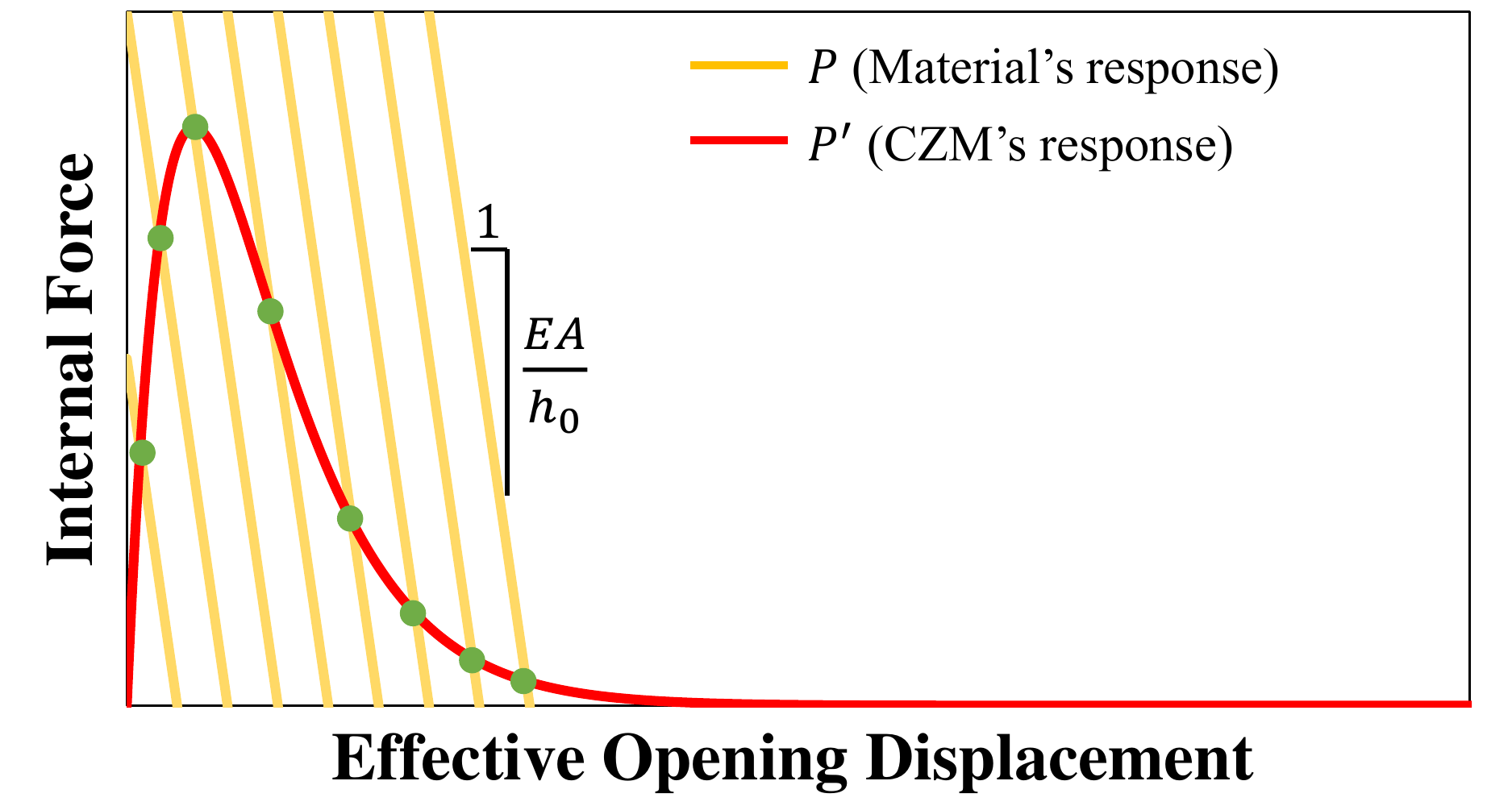}
        \caption{} \label{fig:2b}
    \end{subfigure}
    \\
    \begin{subfigure}{0.65\textwidth}
	    \centering
        \includegraphics[width=\textwidth]{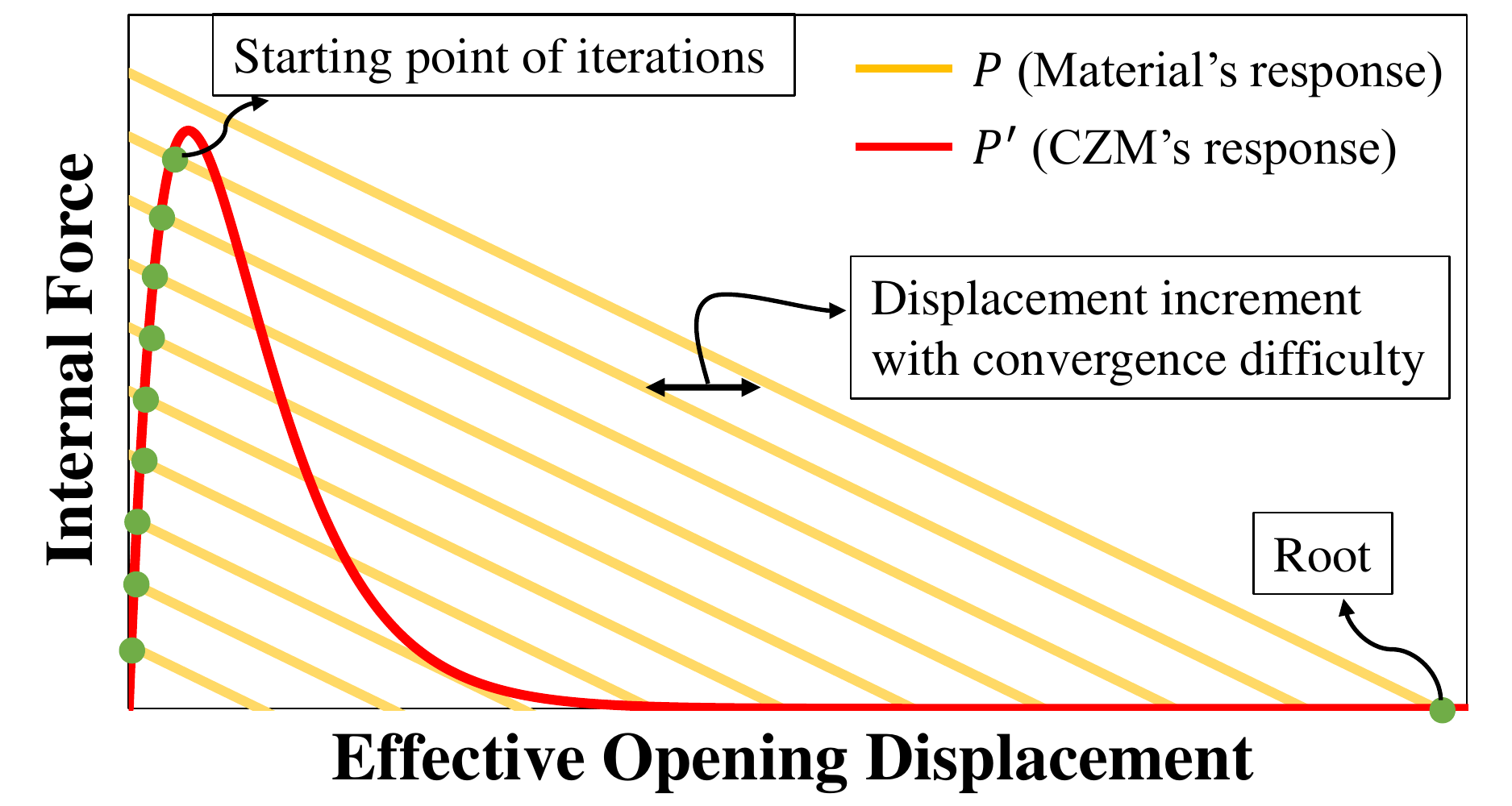} 
        \caption{} \label{fig:2c}
    \end{subfigure}

	\caption{(a) A simple model with a horizontal interface between two elastic materials, and (b) and (c) represents the simple model responses for the cases when the material stiffness (i.e., $EA/h_0$) is close to the CZM’s local tangent stiffness, and when the material stiffness is less than the CZM's local tangent stiffness, respectively. The green dots represent equilibrium points, which are the intersection of the elastic material and CZM responses, under monotonically increasing applied displacement.}
	\label{FIG:NR iterations for stiff anf soft EE}
\end{figure}

To demonstrate the simulation of FE models with cohesive interfaces under static displacement controlled conditions, a simple model, as presented in \textcolor{blue}{Fig.~\ref{FIG:NR iterations for stiff anf soft EE}~(a)}, is considered. The FE model consists of two identical elastic 4-node elements with a modulus of elasticity $E$ that are connected via a cohesive interface. The elastic elements have a length of $h_0$, and a cross-sectional area of $A$. The displacement, $\Delta$, is applied to the top of the model and is increased monotonically. The goal is to calculate the vertical displacement of the free nodes such that equilibrium is satisfied. 
Each free node of the interface element is shared by an adjacent elastic element, as shown with red dots in \textcolor{blue}{Fig.~\ref{FIG:NR iterations for stiff anf soft EE}~(a)}.
For equilibrium to be maintained throughout the analysis, the sum of the internal forces at each interfacial node must be equal to zero. The internal forces include the forces applied by the material ($P'$) and the CZM ($P$). Therefore, at each displacement increment, $P'-P= 0$ must hold. 

\textcolor{blue}{Fig.~\ref{FIG:NR iterations for stiff anf soft EE}~(b) and (c)} illustrate the solution to the described example. The red line represents the CZM's response, while the yellow lines are material's responses to the applied displacement. The horizontal distances between the yellow lines are equal to the applied displacement increments. At each applied displacement increment, equilibrium is satisfied where the material's and CZM's responses intersect. The equilibrium points, as shown with green dots in \textcolor{blue}{Fig.~\ref{FIG:NR iterations for stiff anf soft EE}~(b) and (c)}, need to be computed at each displacement increment using the NR method. It is evident in \textcolor{blue}{Fig.~\ref{FIG:NR iterations for stiff anf soft EE}~(b)} that when the material stiffness (i.e., $EA/h_0$) is larger than the CZM's local tangent stiffness, the equilibrium points are close to one another. Hence, the convergence is easy and fast throughout the analysis. However, when the material stiffness is less than the CZM's local tangent stiffness, the solution jump occurs at the CZM's failure increment, as seen in \textcolor{blue}{Fig.~\ref{FIG:NR iterations for stiff anf soft EE}~(c)}. As a consequence, the starting point of iterations in the NR method and the root are far apart, and the convergence is difficult.

\begin{figure}
	\centering
		\includegraphics[height=2.1 in]{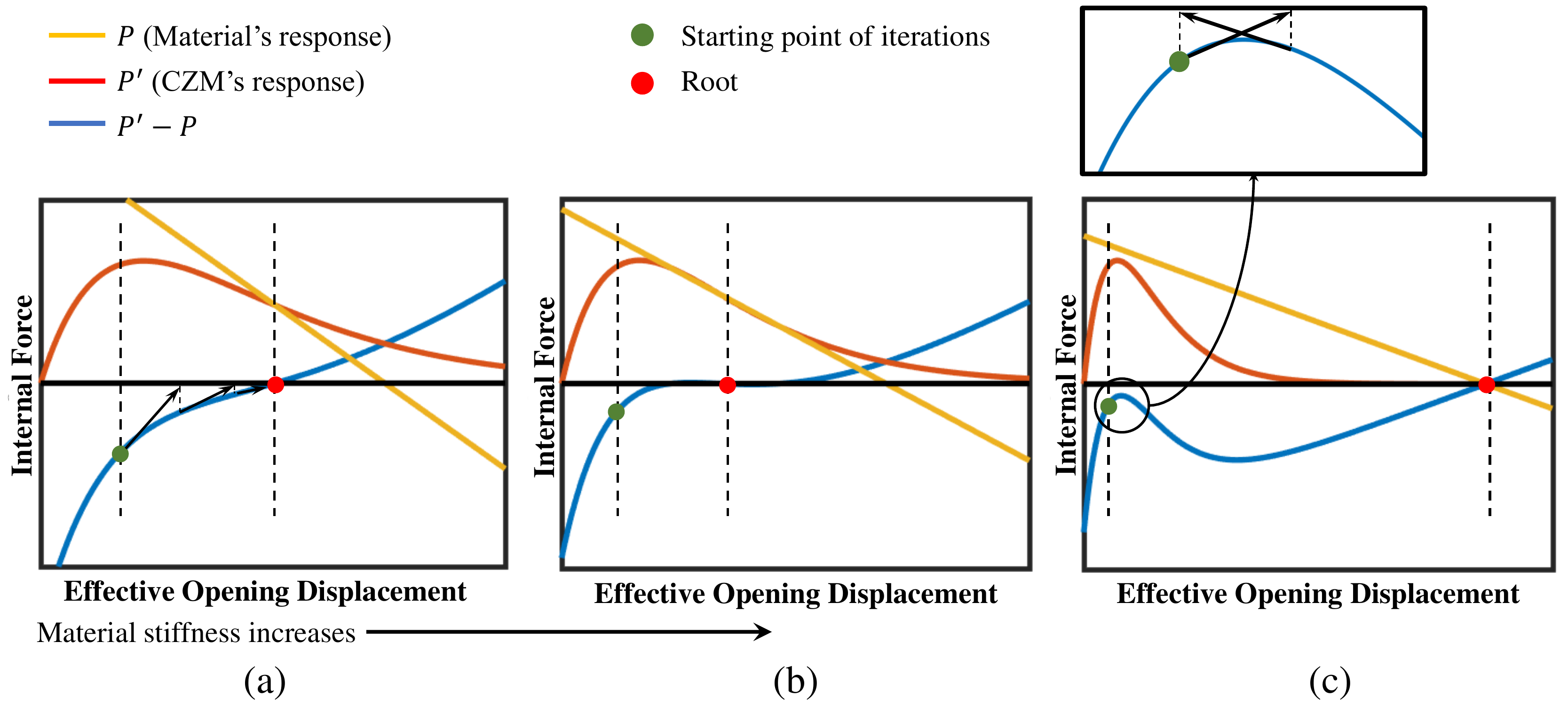}
	\caption{The solutions of the simple elastic bar model at the cohesive zone failure step for the cases of (a) an easy convergence, (b) the threshold between an easy and difficult convergence, and (c) a difficult convergence.}
	\label{FIG:NR iterations as the material stiffness varies}
\end{figure}

It is worthwhile to study the NR iterations at the CZM's failure increment more closely to examine the effects of the solution jump and the relative stiffness. The NR step to solving for $P-P'=0$ at the failing of cohesive zone increment is illustrated in \textcolor{blue}{Fig.~\ref{FIG:NR iterations as the material stiffness varies}} for three different material stiffness. It must be noted that the previous converged solution is always somewhere around CZM's failure point. The blue curves in the plots represent $P-P'$. Hence, the root is where this curve intersects with the horizontal axis. The different cases presented in \textcolor{blue}{Fig.~\ref{FIG:NR iterations as the material stiffness varies}} are as follows:
\begin{enumerate}
\item[(a)] The material is stiff relative to the CZM's local tangent stiffness. Therefore, the $P-P'$ has a well-behaved ascending function, and the NR iterations converge to its root with no difficulty, as it is schematically illustrated in \textcolor{blue}{Fig.~\ref{FIG:NR iterations as the material stiffness varies}~(a)}.
\item[(b)] The material's stiffness is equal to the CZM's local tangent stiffness. Hence, the resulting $P-P'$ function has an inflection point where it intersects with the horizontal axis, as shown in \textcolor{blue}{Fig.~\ref{FIG:NR iterations as the material stiffness varies}~(b)}. This case forms a threshold between an easy and a difficult convergence. 
\item[(c)] The material is soft relative to the CZM's local tangent stiffness. In this case, $P-P'$ function has a local maximum close to the starting point of iterations, which is the last converged solution, as it is schematically shown in \textcolor{blue}{Fig.~\ref{FIG:NR iterations as the material stiffness varies}~(c)}. As a consequence, there is a chance that the NR iterations enter an infinite cycle of iterations and do not converge.
\end{enumerate}

Therefore, it can be seen that in static analyses, numerical instability is the root of the CZMs' convergence difficulty. In light of this knowledge, the performance of the two most popular methods described in Section \ref{subsec: existing methods} (i.e., viscous regularization and decrease of global displacement) can be studied in more detail. 

\subsection{Evaluation of the Two Common Methods to Resolve the Convergence Difficulty: Viscous Regularization, and Decrease of Global Displacement Methods} \label{subsec: Evaluation of existings}
In the viscous regularization method, the inclusion of a rate-dependent term in the cohesive function causes a decrease in the post-failure decay rate. This treatment, in turn, regularizes the local maximum of the blue curve in \textcolor{blue}{Fig.~\ref{FIG:NR iterations as the material stiffness varies}~(c)} and forces it towards an inflection point with a horizontal slope similar to the case in \textcolor{blue}{Fig.~\ref{FIG:NR iterations as the material stiffness varies}~(b)}. The regularization facilitates the convergence. However, as a consequence, an instantaneous failure is simulated as being slow and non-instantaneous, which may significantly affect the stress distribution within the FE analysis. Therefore, a smaller viscosity coefficient needs to be used to ensure that the viscosity is not negatively affecting the stress distribution. A smaller viscosity coefficient leads to several time step refinements.

\begin{figure}
	\centering
		\includegraphics[height=2.23 in]{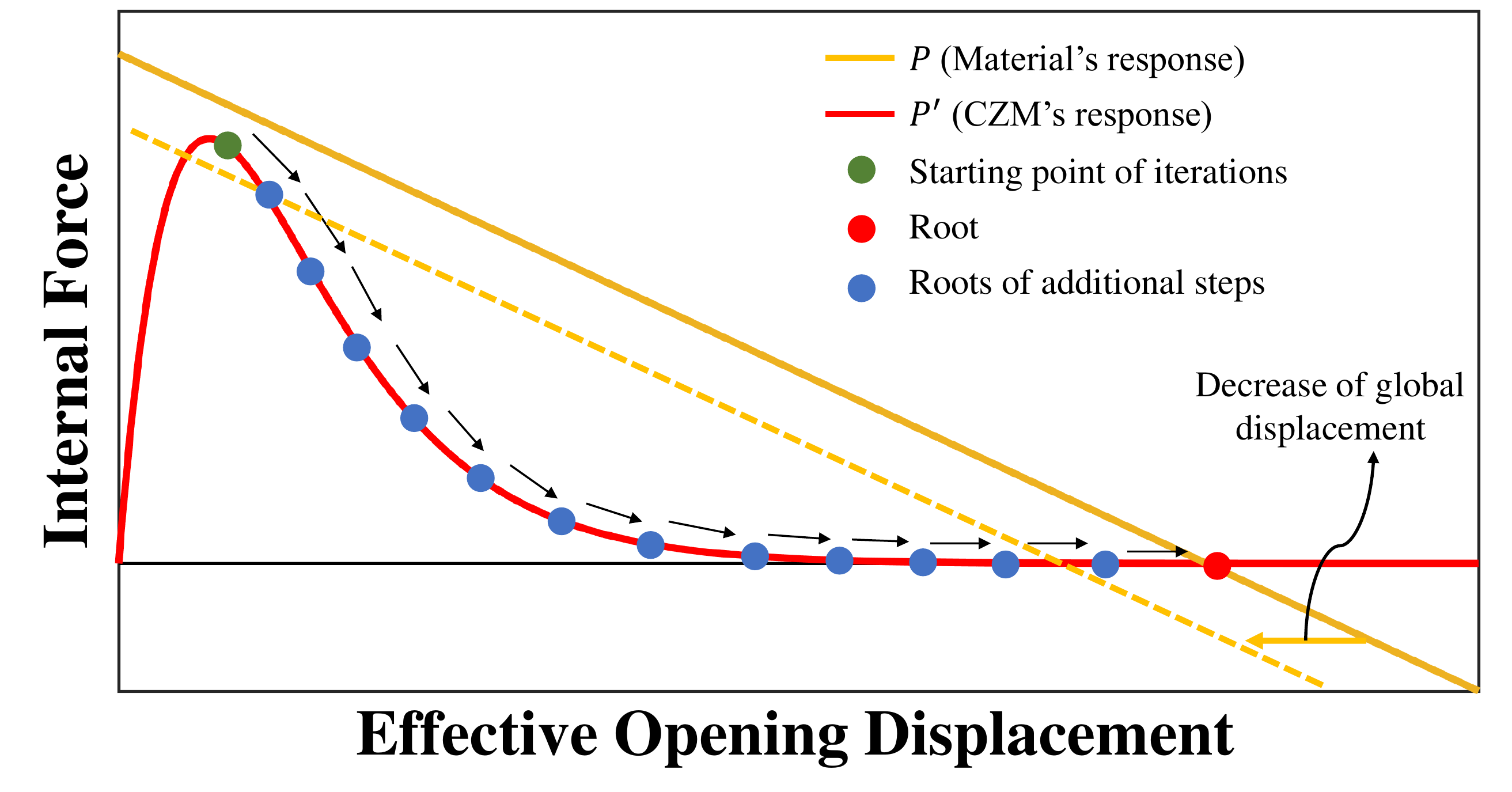}
	\caption{The path to be traveled towards convergence via extra displacement increments in the decrease of global displacement method.}
	\label{FIG:decrease of global displacement}
\end{figure}

In the decrease of global displacement method, the applied displacement is decreased after the convergence difficulty is encountered. As a result, the convergence is towards the roots on the decaying portion of the CZM through additional equilibrium points, as illustrated with blue dots in \textcolor{blue}{Fig.~\ref{FIG:decrease of global displacement}}. 
As can be seen, the convergence is ultimately towards the root that could not converge in the first place (shown with a red dot). As a result, the method converges to the accurate root eventually and can simulate an instantaneous failure. However, the method is complex to implement and is computationally expensive. 

\section{Proposed Novel Method to Resolve CZMs' Convergence Difficulty} \label{proposed approach}
The previous section described that an inappropriate starting point of iterations, located in the vicinity of a local maximum, is responsible for the convergence difficulty of CZMs. This section presents a method for addressing the convergence issue. The proposed procedure is based on modifying the starting point of the problematic NR step at the failure increment. The modification is applied only to the pair nodes of the interface elements (PNIEs), causing the convergence problem. The term ``starting point of iterations" in this context refers to the effective opening displacement of PNIEs at the previous converged step.

\begin{figure}
	\centering
	\begin{subfigure}{1\textwidth}
	    \centering
        \includegraphics[width=4.7 in]{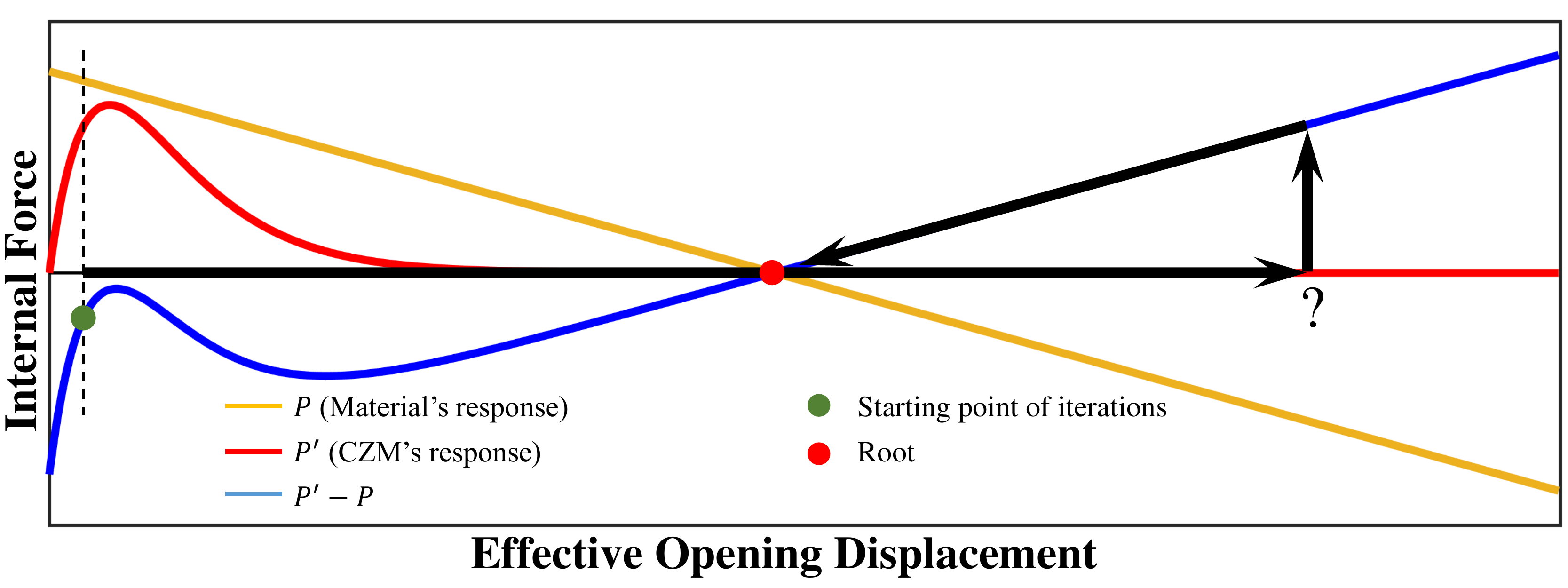}
       \caption{} 
    \end{subfigure}
	\begin{subfigure}{1\textwidth}
	    \centering
        \includegraphics[width=4.7 in]{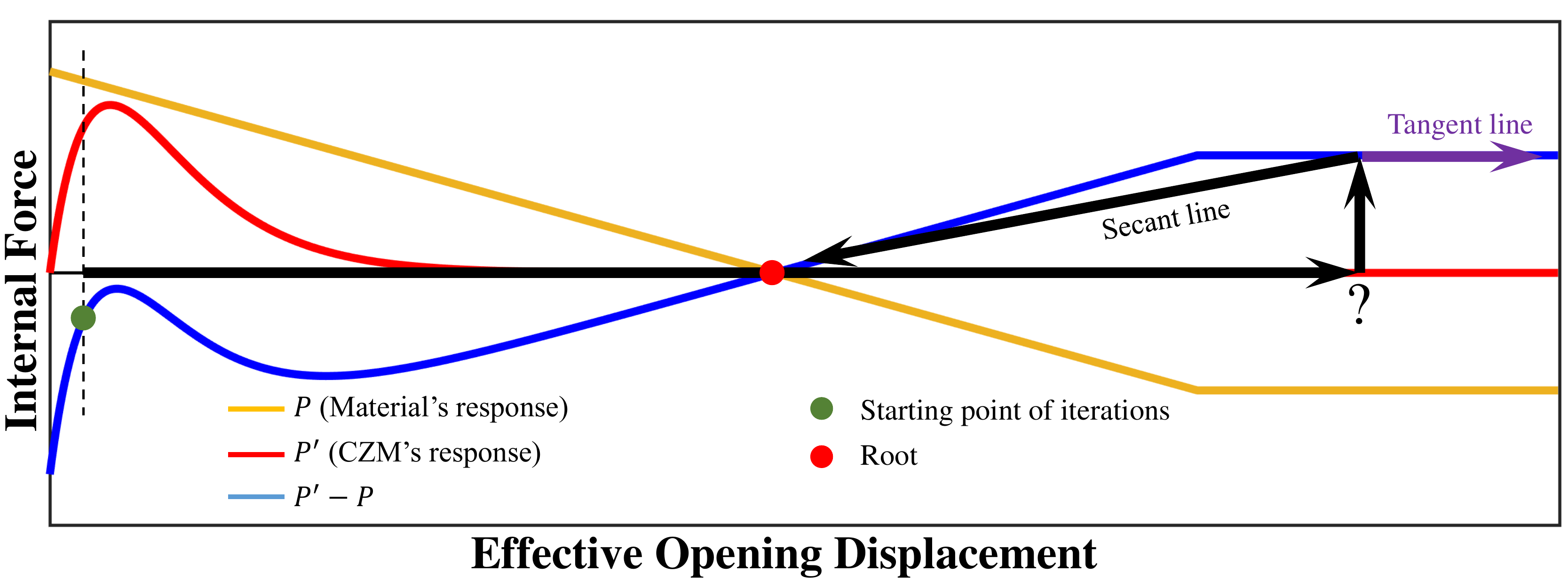}
      \caption{}  
    \end{subfigure}    
	\caption{Modification of the starting point of iterations for the case of (a) elastic material, and (b) nonlinear material behavior, which can be defined by either tangent stiffness or secant stiffness.}
	\label{FIG:modification with secant}
\end{figure}

\textcolor{blue}{Fig.~\ref{FIG:modification with secant}~(a)} illustrates the solution to the simple model presented in \textcolor{blue}{Fig.~\ref{FIG:NR iterations for stiff anf soft EE}~(a)} at the interface's failure increment for a case with convergence difficulty. It is noted that if the starting point of iterations locates anywhere on the right linear ascending portion of the $P-P'$ function, the NR iterations quickly converge to the root. Hence, the displacement of the problematic PNIE can be modified such that the starting point of iterations occurs on the linear ascending portion of the $P-P'$ function. For a linear elastic material behavior, any starting point on the linear ascending part leads to fast and easy convergence.

\begin{figure}
	\centering
		\includegraphics[height=2 in]{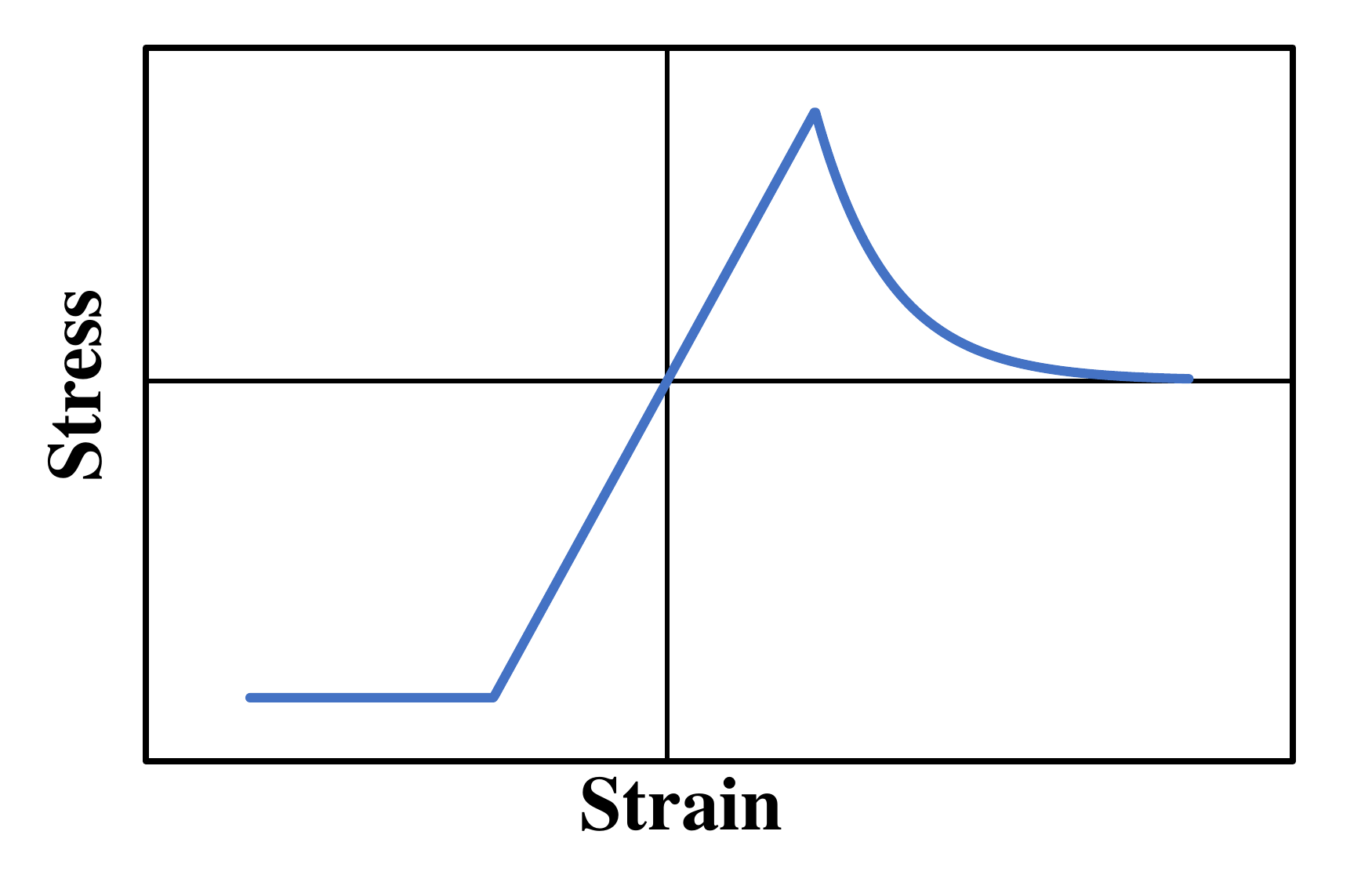}
	\caption{An example of a nonlinear behavior for a brittle material.}
	\label{FIG:nonlinear material}
\end{figure}

For the case of materials with nonlinear behavior, a constitutive law as presented in \textcolor{blue}{Fig.~\ref{FIG:nonlinear material}} is assumed as a typical example. The material is brittle under tension and elastic-perfectly-plastic under compression.
For the simple model presented by \textcolor{blue}{Fig.~\ref{FIG:NR iterations for stiff anf soft EE}~(a)}, any enforced debonding along the debonding direction subjects the material to pure compression.
As a result, the rightmost portion of the $P$ function, as illustrated in \textcolor{blue}{Fig.~\ref{FIG:modification with secant}~(b)}, becomes horizontal following the material's behavior under compression (i.e., elastic-perfectly-plastic). 
The leftmost portion of the $P'-P$ function follows the material's behavior under tension, which is not included in the plot as it does not affect the proposed method. When the modification is applied to the starting point of iterations, it may end up on the horizontal portion of $P'-P$ function, as illustrated in \textcolor{blue}{Fig.~\ref{FIG:modification with secant}~(b)}. Once the starting point locates on the horizontal part, the performance of the proposed method in addressing the convergence issue depends on how the material's stiffness matrix is defined in an FE framework. 

If the material's behavior is defined using the material's tangent stiffness matrix, once increasing the starting point of iterations, the NR diverges from the root following the tangent slope of the $P'-P$ function. In FE simulations, however, the secant stiffness matrix is typically used to define a nonlinear material behavior, particularly when damage models represent the nonlinearity. If the secant stiffness matrix defines the material's constitutive behavior, the iterations quickly converge to the root following the secant slopes of the material's response, as it can be seen in \textcolor{blue}{Fig.~\ref{FIG:modification with secant}~(b)}. Hence, the starting point can be increased by any large magnitude to ensure that it locates beyond the problematic local maximum of the $P'-P$ function, and the convergence is guaranteed. 

When the secant stiffness is used to define the material's behavior, whereas the tangent stiffness defines the CZM's response, the following features can be mentioned regarding the global stiffness matrix:
\begin{enumerate}
    \item[(a)] At the initial nonlinear portion of the $P'-P$ function in \textcolor{blue}{Fig.~\ref{FIG:modification with secant} (b)}, the iterations follow a path defined by the tangents of the blue curve since the CZM's nonlinear behavior is determined by tangent stiffness (i.e., the red line).
    \item[(b)] At the central linear ascending portion of the $P'-P$ function where the contribution of CZM's behavior is diminished, the tangent and secant stiffness matrices are the same.
    \item[(c)] At the end horizontal portion of the $P'-P$ function, the iterations follow a path defined by the material's secant stiffness.
\end{enumerate}

\begin{figure}
	\centering
		\includegraphics[height=2 in]{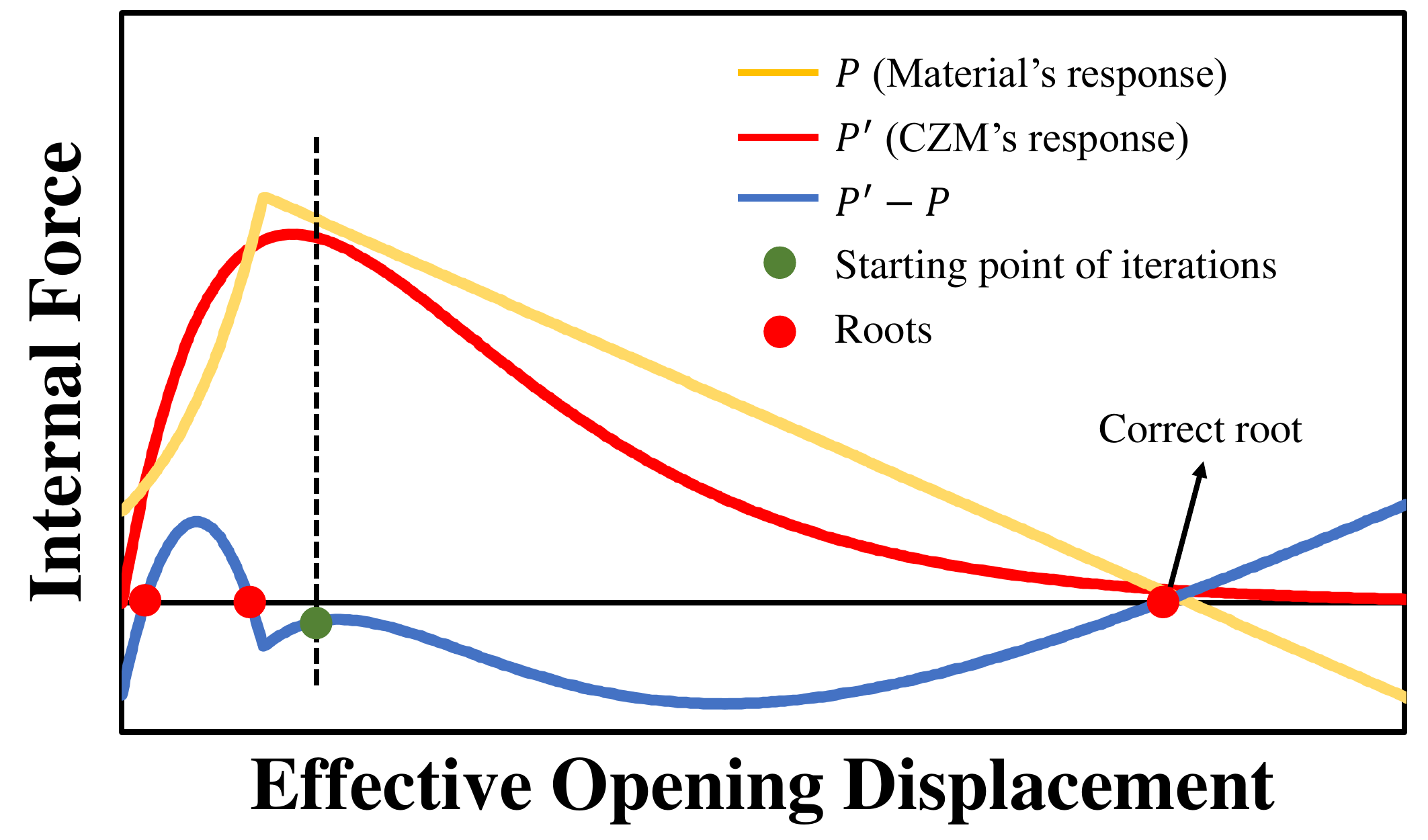}
	\caption{Presence of multiple roots for the step with convergence difficulty if the material is brittle under tension with a steep post-failure decay.}
	\label{FIG:effect of material nonlinearity under tension}
\end{figure}

The effect of the material's tensile behavior on the application of the proposed method is also worth studying. First, if the material's tensile strength is smaller than the CZM's cohesive strength, the material fails before the interface, and the opening displacement of the interface recovers. Since the interface does not fail in such cases, there is no convergence issue resulting from the CZM. Second, if the material's tensile strength is greater than the CZM's cohesive strength, the interface fails, and the CZM might cause convergence difficulty. In this case, if the material is brittle with a steep post-failure decay in tension, there may be three equilibrium roots at the failure increment, as it is depicted in \textcolor{blue}{Fig.~\ref{FIG:effect of material nonlinearity under tension}}. The correct root is the one that corresponds to the failure of the interface, followed by a deformation recovery in the material (the rightmost root). Since the modification increases the abscissa of the starting point of iterations, the convergence is still towards the rightmost root in \textcolor{blue}{Fig.~\ref{FIG:effect of material nonlinearity under tension}}. As a result, the presence of the other two roots has no influence on the performance of the proposed method.

Thus far, the performance of the proposed method was studied on a simple model. In general, interfaces may undergo mixed-modes of debonding in models with complex geometries. Hence, the application of the proposed modification to a PNIE causes a general state of stress to be induced within the adjacent material. Under this condition, the response of the material (i.e., the shape of the yellow curve in \textcolor{blue}{Fig.~\ref{FIG:modification with secant} (b)}) is tough to predict. However, as long as the secant stiffness defines the material's constitutive behavior, the proposed method does not depend on the shape of the material's response. Therefore, the presented discussions can be generalized to any arbitrary mode of debonding. Hence, any problematic PNIE can be simply modified as described in this section once they are detected in the FE model. The algorithm for detecting and modifying the problematic PNIEs are explained in detail in the following subsections.

\subsection{Detecting the problematic PNIEs}
The process of detecting problematic PNIEs for the case of CZMs with the exponential behavior as presented in \textcolor{blue}{Eqn.~\ref{eq:1}} is explained in this section. Detecting the problematic PNIEs for CZMs with different functions can be achieved following the same procedure. 

As it was explained in the previous section, the oscillation during the infinite cycle occurs about the local maximum of the $P-P'$ function, as it can be seen in \textcolor{blue}{Fig.~\ref{FIG:NR iterations as the material stiffness varies} (c)}. Hence, one way of detecting the problematic PNIEs is to calculate the abscissa of the $P-P'$ function's local maximum for all PNIEs in the FE model. The next step is to check if the effective opening displacement of the PNIE is oscillating about the local maximum point. This process, however, requires the knowledge of the material's stiffness (i.e., slope of the yellow line in \textcolor{blue}{Fig.~\ref{FIG:NR iterations as the material stiffness varies} (c)}) corresponding to all PNIEs throughout the FE model, which is costly to compute. An easier and more efficient way of detecting the problematic PNIEs can be achieved. It can be shown that the starting point of iterations corresponding to a problematic PNIE, always occurs within an interval.

\begin{figure}
	\centering
		\includegraphics[width=4.7 in]{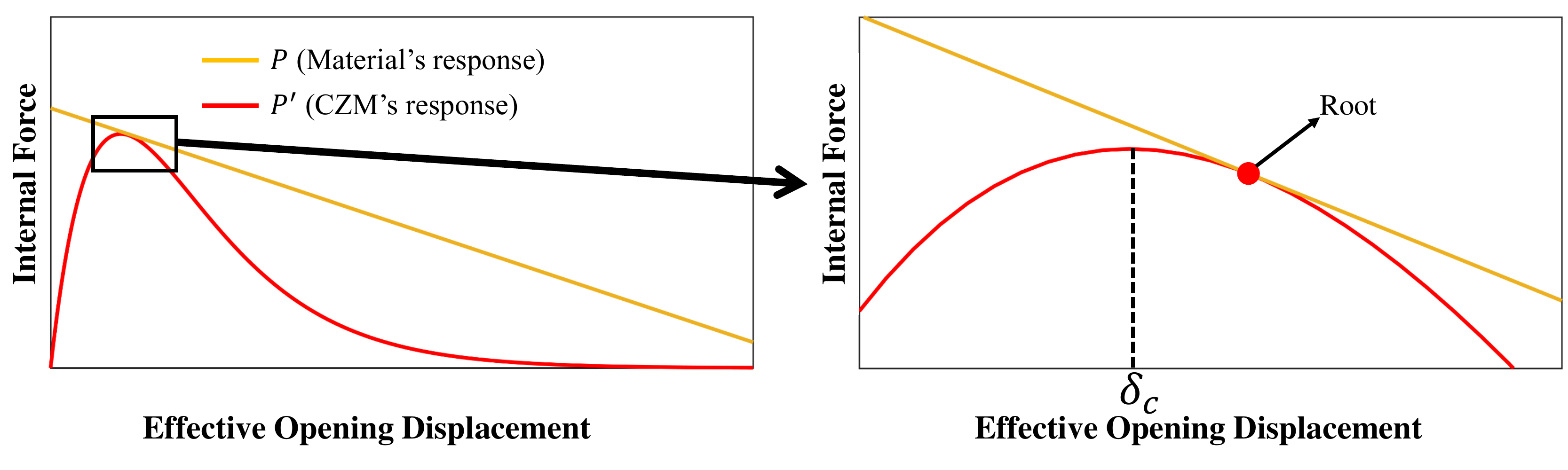}
	\caption{Location of starting point of iterations for the step with convergence difficulty.}
	\label{FIG:starting point of iterations for the problematic PNIEs}
\end{figure}

The lower bound of the interval can be calculated by realizing the fact that the starting point of iterations for the problematic PNIEs is always greater than $\delta_c$.
This is schematically illustrated in \textcolor{blue}{Fig.~\ref{FIG:starting point of iterations for the problematic PNIEs}}, which depicts the converged solution of the previous step (i.e., the starting point of iterations for the current step).
As a result, the lower bound of the interval is equal to $\delta_c$.

The upper bound of the interval corresponds to the threshold case between an easy and a difficult convergence, as it can be seen in \textcolor{blue}{Fig.~\ref{FIG:NR iterations as the material stiffness varies}~(b)}. 
In this case, neglecting the Poisson's ratio effect, $P$ and $P'$ can be defined as follows:
\begin{eqnarray}\label{eq:P'}
P'=\frac{\sigma_c}{\delta_c}\delta \; e^{(1-\delta/\delta_c)}(\frac{A}{2})
\end{eqnarray}
\begin{eqnarray}\label{eq:P}
P=B_1 \delta-B_0
\end{eqnarray}
where $B_1$ and $B_0$ are two unknowns. $A/2$ indicates that half of the load in the interface is carried by each PNIE. The $P'-P$ in the vicinity of the root for the previous converged step is
\begin{eqnarray}\label{eq:4}
P'-P=\frac{\sigma_c}{\delta_c}\delta \; e^{(1-\delta/\delta_c)}(\frac{A}{2})-(B_1 \delta-B_0)
\end{eqnarray}

The root in the threshold case is the abscissa of the inflection point, which can be calculated by taking the second derivative of \textcolor{blue}{Eqn.~\ref{eq:4}} and setting it equal to zero, which is
\begin{eqnarray}\label{eq:5}
\frac{d^2(P'-P)}{d\delta^2}=\sigma_ce^{(1-\delta/\delta_c)}(\frac{A}{2})\left(\frac{-2}{\delta_c^2}+\frac{\delta}{\delta_c^3}\right)=0
\end{eqnarray}
By solving the above equation, the interval's upper bound is calculated to be $\delta=2\delta_c$.

Hence, the effective opening displacements of the problematic PNIEs at the beginning of the iterations for the step with convergence difficulty always occurs somewhere between $\delta_c$ and $2\delta_c$. It must be noted that the calculated interval is typically very short as $\delta_c$ is normally chosen to be a very small number to ensure strain continuity before the failure of the interface. However, there is still a small chance that the effective opening displacement of a non-problematic PNIE occurs within the interval at the moment of instability. In this case, the $P-P'$ is a monotonically increasing function, as illustrated in \textcolor{blue}{Fig.~\ref{FIG:NR iterations as the material stiffness varies}~(a)}, and hence, any starting point of iterations has an easy convergence. Therefore, applying the modification to a non-problematic PNIE does not cause any issue. 

\subsection{Modification of the problematic PNIEs}
\begin{figure}
	\centering
	\begin{subfigure}{0.49\textwidth}
	    \centering
        \includegraphics[height=2.5 in]{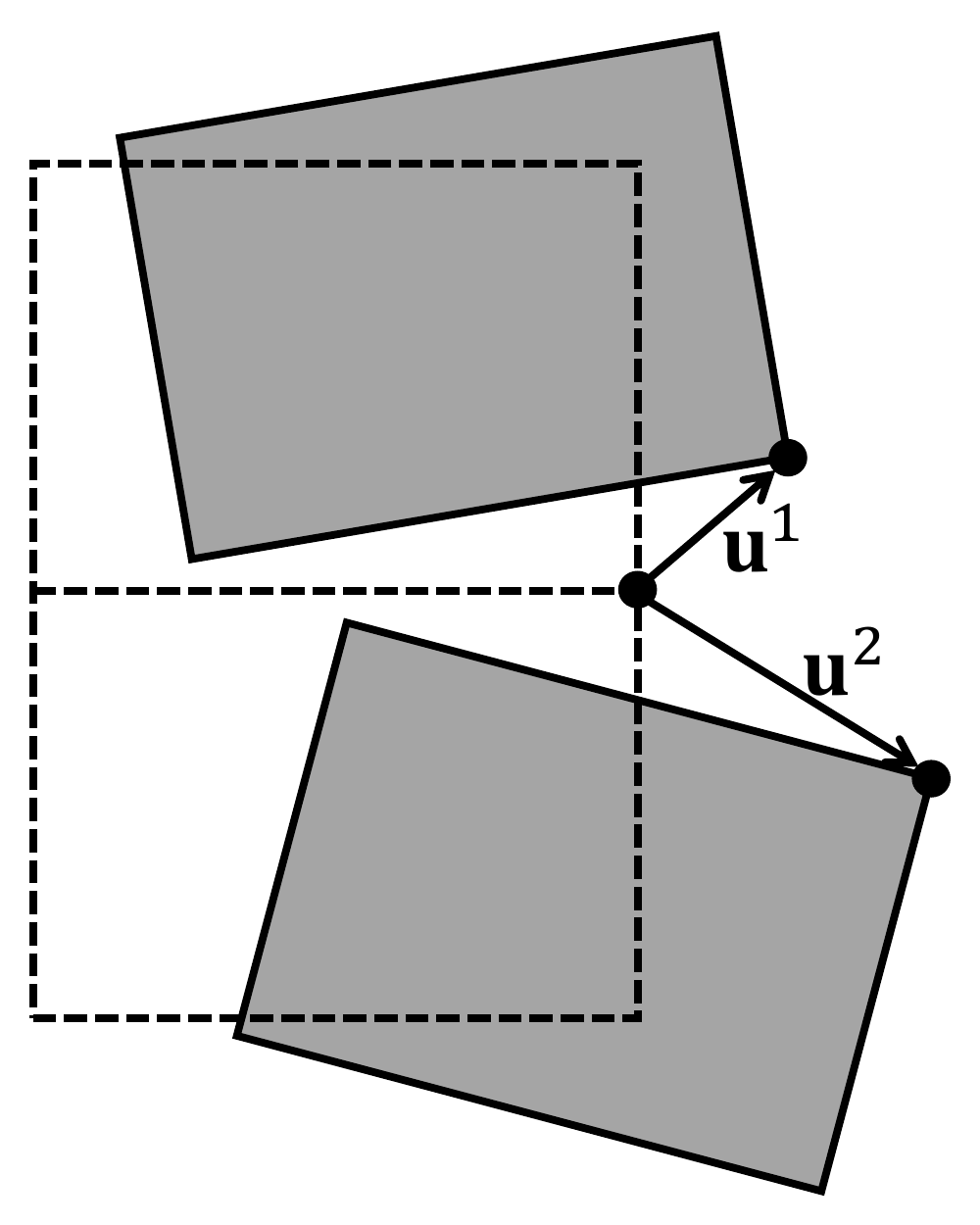}
       \caption{} \label{FIG:9a}
    \end{subfigure}
	\begin{subfigure}{0.49\textwidth}
	    \centering
        \includegraphics[height=2.5 in]{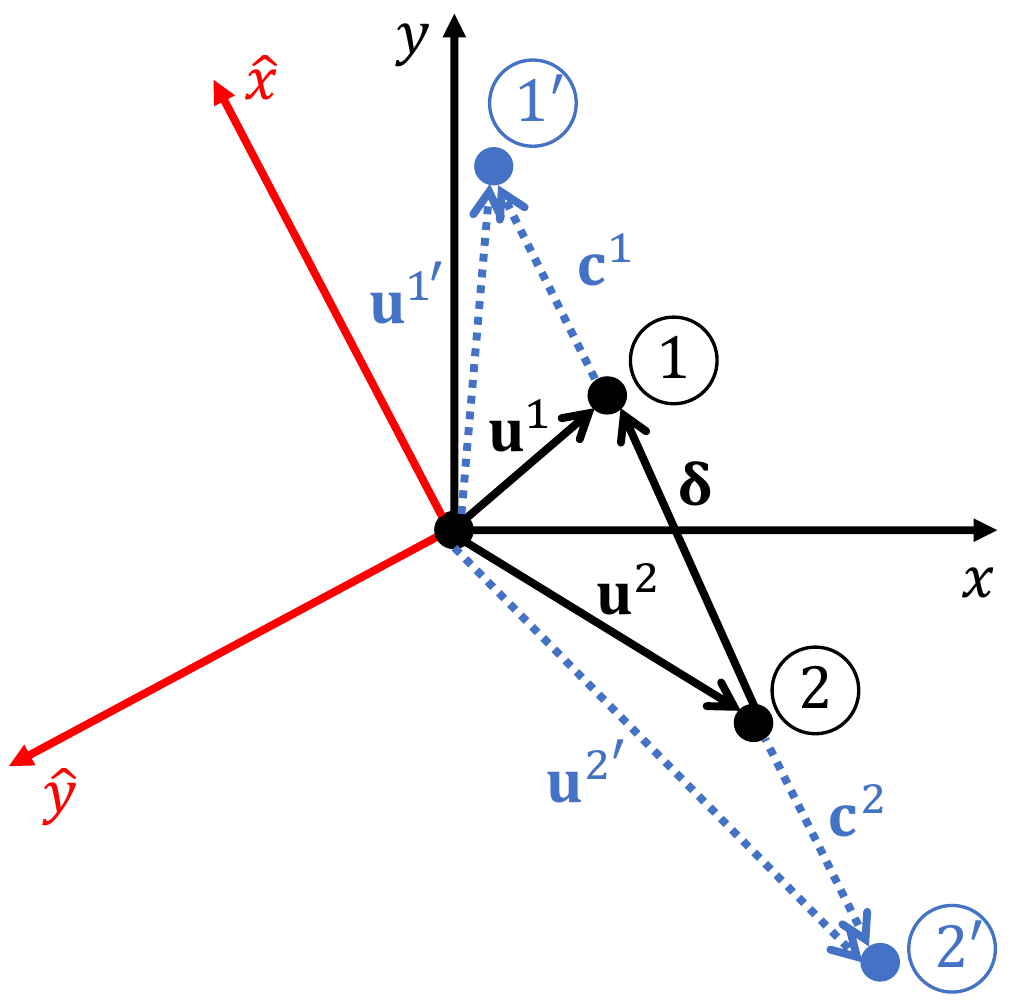}
      \caption{}  \label{FIG:9b}
    \end{subfigure}    
	\caption{(a) Schematic representation of an arbitrary deformation of two rectangular elements connecting via an interface element, and (b) the modification applied to the PNIEs' displacement vectors.}
	\label{FIG:PNIE displacement correction}
\end{figure}

Once the problematic PNIEs are detected, the next step is to modify them to a proper starting point of iterations. \textcolor{blue}{Fig.~\ref{FIG:PNIE displacement correction}~(a)} illustrates an arbitrary deformation of a cohesive interface with a problematic PNIE. The nodal displacement vectors $\mathbf u^{\,1}$ and $\mathbf u^{\,2}$ can be extracted directly from the displacement vector of the previous converged step in the global $x - y$ coordinate system.
The goal is to calculate $\mathbf {u}^{\,1'}$ and $\mathbf {u}^{\,2'}$ through vectors $\mathbf {c}^{\,1}$ and $\mathbf {c}^{\,2}$ such that:
\begin{enumerate}
    \item[(a)] the resulting opening displacement vector of the PNIE (i.e., $\boldsymbol{\updelta'}=\mathbf {u}^{\,1'}-\mathbf {u}^{\,2'}$) is aligned with the opening displacement vector before the modification (i.e., $\boldsymbol{\updelta}=\mathbf {u}^{\,1}-\mathbf {u}^{\,2}$), as it is shown in \textcolor{blue}{Fig.~\ref{FIG:PNIE displacement correction}~(b)}.
    \item[(b)] $\boldsymbol{\updelta'}$ has a much greater magnitude than $\boldsymbol{\updelta}$.
\end{enumerate}
Vectors $\mathbf {c}^{\,1}$ and $\mathbf {c}^{\,2}$ both have arbitrary large magnitudes, and for simplicity, they can be assumed to be equal (i.e., $|\mathbf {c}^{\,1}|=|\mathbf {c}^{\,2}|=c$). 

To apply the modification, the vectors $\mathbf {u}^{\,1}$ and $\mathbf {u}^{\,2}$ are first transformed into another orthonormal basis $\hat{x}$ - $\hat{y}$ such that $\hat{x}$ is aligned with $\boldsymbol{\updelta}$. The transformation can be performed through a transformation matrix $\mathbf T$ as
\begin{equation} \label{eq:7}
\begin{array}{cc} \mathbf {\hat{u}}^{\,1}=\mathbf T^{-1}\mathbf {u}^{\,1}=\mathbf T^T\mathbf {u}^{\,1} \\
\mathbf {\hat{u}}^{\,2}=\mathbf T^{-1}\mathbf {u}^{\,2}=\mathbf T^T\mathbf {u}^{\,2} \end{array}
\end{equation}
Now the $\hat{x}$ component of the displacements can be modified as
\begin{eqnarray}\label{eq:8}
\begin{array}{cc}  {\hat{u}}^{\,1'}_{\hat{x}}= {\hat{u}}^{\,1}_{\hat{x}}+c\\
{\hat{u}}^{\,2'}_{\hat{x}}={\hat{u}}^{\,2}_{\hat{x}}-c\end{array}
\end{eqnarray}
After the displacements are modified, the new displacement vectors can be transformed back to the $x - y$ coordinate system as
\begin{eqnarray}\label{eq:9}
\begin{array}{cc} \mathbf {u}^{\,1'}=\mathbf T\mathbf {\hat{u}}^{\,1}\\
\mathbf {u}^{\,2'}=\mathbf T\mathbf {\hat{u}}^{\,2}\end{array}
\end{eqnarray}
$\mathbf {u}^{\,1'}$ and $\mathbf {u}^{\,2'}$ are the modified values and must be used instead of the old values of $\mathbf {u}^{\,1}$ and $\mathbf {u}^{\,2}$ inside the displacement vector at the beginning of the iterations.

\section{Verification and Application Examples}
To evaluate the performance of the proposed method, an FE program was developed in Matlab for 2-D plane strain analyses, which includes cohesive element interfaces. The coupling scenario presented by \textcolor{blue}{Eqn.~\ref{eq:2}} is used for the interfaces, with $\beta$ assumed to be 1. Three application examples, including a 2-D bar with a horizontal interface under tension, a double-cantilever beam, and a 2-D bar with a circular interface under tension, are presented in this section. The latter example is used to verify the functionality of the proposed method in mixed-mode debonding.
The simulations results using the proposed modification method are compared with Abaqus-implicit \textcolor{blue}{\cite{abaqus}} and viscous regularization method.

The viscous regularization method proposed by Gao and Bower \textcolor{blue}{\cite{gao2004simple}} is used in the analyses, which introduces the addition of a rate-dependent term to the cohesive law as
\begin{eqnarray}\label{eq:11}
\sigma_{visc}=\xi\frac{\sigma_c}{\delta_c}\left(\frac{d\delta}{dt}\right)
\end{eqnarray}
where $\sigma_{visc}$ is the artificial viscosity that is added to the cohesive function, $\xi$ is the viscosity parameter, and $t$ is the analysis pseudo-time. To minimize the effect of the viscous regularization method on the accuracy, $\xi$ is chosen to be the smallest value that prevents the termination of the analysis due to convergence difficulty. 

For all the application examples, the material behavior is elastic under tension and elastic-perfectly-plastic under compression defined by secant stiffness, with a yield strength of $\sigma_y^{comp}$. The nonlinear compressive behavior is to verify that the material's nonlinear behavior under compression does not affect the performance of the proposed method, as discussed in Section \ref{proposed approach}.

\subsection{2-D bar with a horizontal interface}
The first example includes a 2-D bar with a cohesive interface. The displacement is applied to the top boundary, while the vertical displacement of the bottom boundary is restrained. The exponential model presented in \textcolor{blue}{Eqn.~\ref{eq:1}} was assumed as the cohesive interface behavior. The properties of the model are presented in \textcolor{blue}{Table \ref{rectangular example properties}}. Three mesh sizes of $1\times1$, $9\times9$, and $100\times100$ were considered to verify that they all yield the same responses. A displacement of $0.2\;mm$ was applied in 100 increments. 

\begin{table}
    \centering
    \caption{Dimensions and material properties for the 2-D bar with a horizontal interface.} \label{rectangular example properties}
\begin{tabular}{|c|c|c|c|c|c|c|}
\hline
\multicolumn{2}{|c|}{\multirow{2}{*}{Dimensions}} & \multicolumn{3}{c|}{\multirow{2}{*}{Material Properties}} & \multicolumn{2}{c|}{Cohesive Properties} \\
\multicolumn{2}{|c|}{}                           & \multicolumn{3}{c|}{}                                     & \multicolumn{2}{c|}{(exponential model)} \\ \hline
Height                  & Width                  & $E$             & $\nu$        & $\sigma_y^{comp}$        & $\sigma_c$          & $\delta_c$         \\
$(mm)$                  & $(mm)$                 & $(MPa)$         &              & $(MPa)$                  & $(MPa)$             & $(mm)$             \\ \hline
1                       & 1                      & 1000            & 0            & 60                       & 60                  & 0.02               \\ \hline
\end{tabular}
\end{table}

\begin{figure}
	\centering
		\includegraphics[height=2 in]{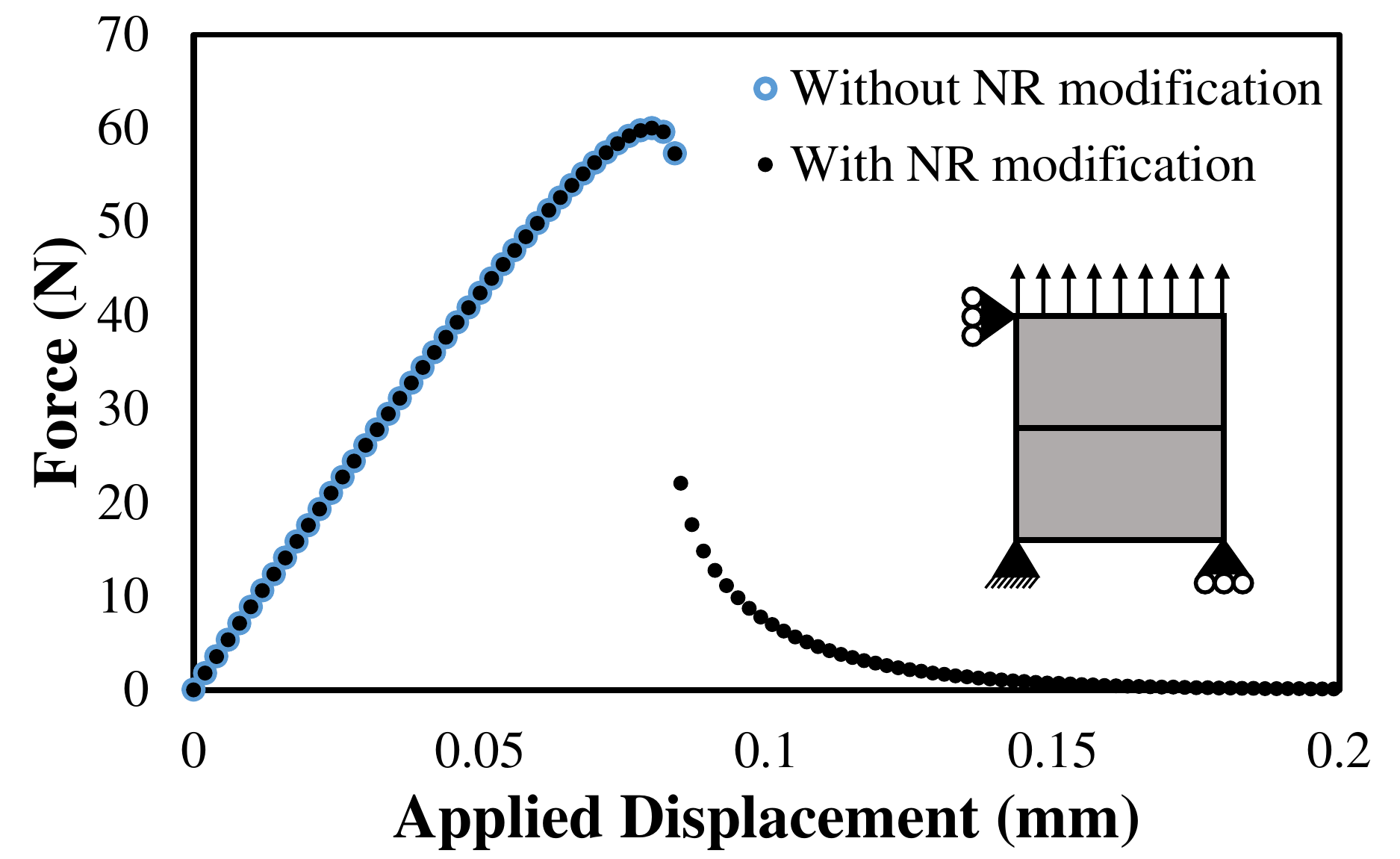}
	\caption{Force-displacement responses for the 2-D bar with a horizontal interface for the cases with and without the application of the NR modification method.}
	\label{FIG:rectangular example response and boundary conditions}
\end{figure}

\textcolor{blue}{Fig.~\ref{FIG:rectangular example response and boundary conditions}} presents the force-displacement responses for the cases with and without the application of the NR modification method. The NR modification was activated only once throughout the analysis, and only three iterations were taken towards convergence of the problematic step. As can be seen in \textcolor{blue}{Fig.~\ref{FIG:rectangular example response and boundary conditions}}, the analysis could not be completed when the proposed modification was not applied. All different meshing scenarios had the exact same responses as expected. The similar results confirm that the modification does not interrupt the system of equations even though several degrees of freedom are involved.

\subsection{Double-cantilever model}
\begin{figure}
	\centering
		\includegraphics[width=4.7 in]{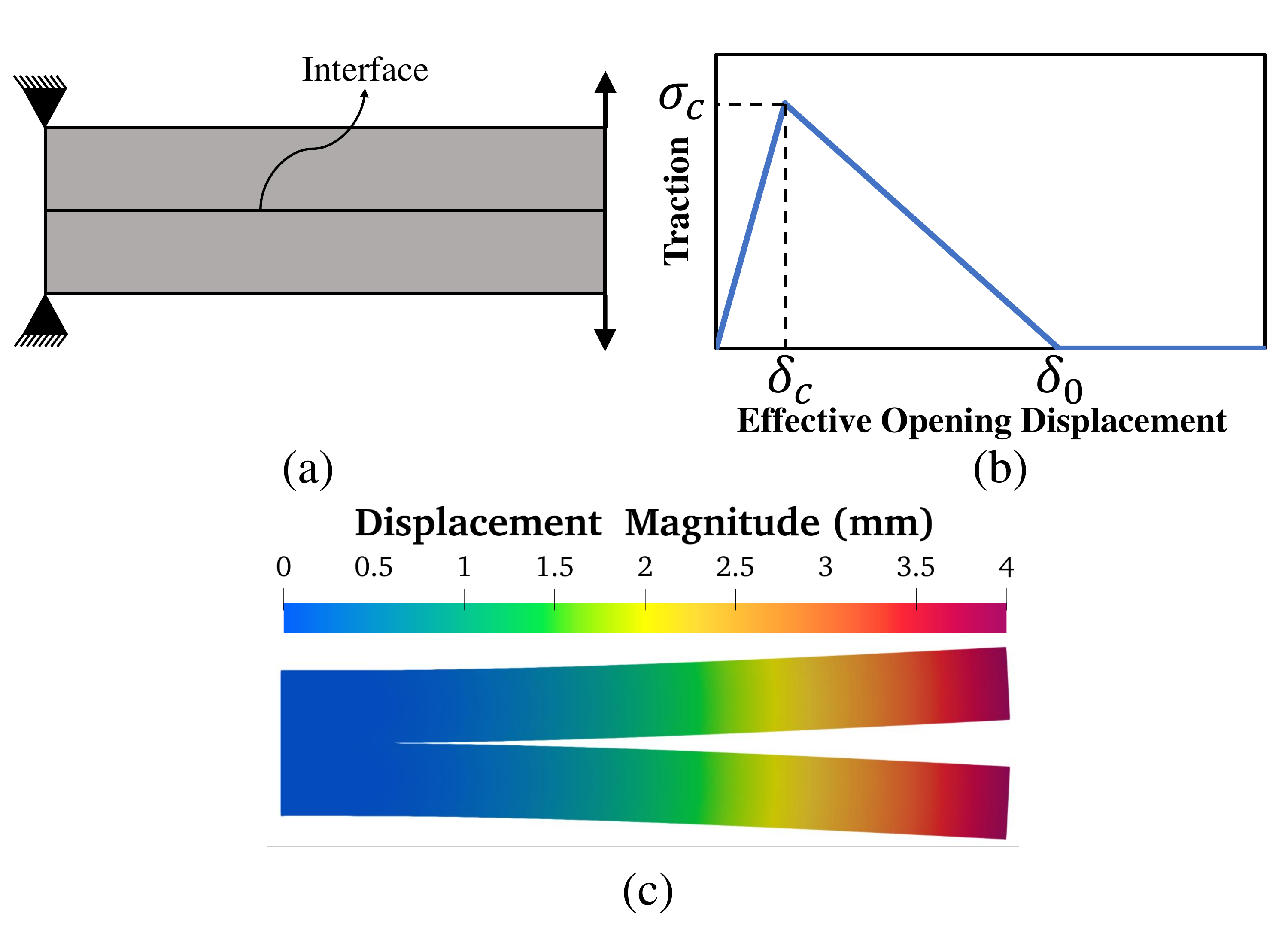}
	\caption{(a) Boundary conditions, (b) bilinear CZM, and (c) deformed shape for the double-cantilever beam model.}
	\label{FIG:double cantilever model}
\end{figure}

The double-cantilever beam is a benchmark example for studying the crack propagation inside materials both experimentally and numerically. \textcolor{blue}{Fig.~\ref{FIG:double cantilever model}~(a)} illustrates the configuration of a double-cantilever beam, and \textcolor{blue}{Table \ref{double cantilever properties}} presents the properties of the model with expected convergence difficulty. The force-displacement response obtained based on the proposed NR modification method is compared with a quasi-static analysis carried out in Abaqus-implicit \textcolor{blue}{\cite{abaqus}} and the viscous regularization approach to validate the accuracy of the proposed method.  It is reminded that the CZMs do not have convergence difficulty in a dynamic analysis. A bilinear CZM was assumed in the analysis, as illustrated in \textcolor{blue}{Fig.~\ref{FIG:double cantilever model}~(b)}. The bilinear model is available in Abaqus and provides the possibility of a precise comparison. The same mesh size and element type were used for all three analyses. A displacement of $4\;mm$ was applied in 400 steps. The displacement was applied very slowly at a rate of $0.004\;mm/s$, the density of the material was set to zero, and a relatively large damping ratio was assumed to minimize dynamic effects in the quasi-static analysis.

\begin{table}[]
    \centering
    \caption{Dimensions and material properties for the double-cantilever beam model.} \label{double cantilever properties}

\begin{tabular}{|c|c|c|c|c|}
\hline
\multicolumn{2}{|c|}{\multirow{2}{*}{Dimensions}} & \multicolumn{3}{c|}{Cohesive Properties}                                   \\
\multicolumn{2}{|c|}{}                            & \multicolumn{3}{c|}{(bilinear model)}                                      \\ \hline
Height                & Width                     & $\sigma_c$                         & $\delta_c$               & $\delta_0$ \\
$(mm)$                & $(mm)$                    & $(MPa)$                            & $(mm)$                   & $(mm)$     \\ \hline
100                   & 500                       & 30                                 & 0.003                    & 0.033      \\ \hline
\multicolumn{5}{|c|}{Material Properties}                                                                                      \\ \hline
\multirow{2}{*}{E}    & \multirow{2}{*}{$\nu$}    & \multirow{2}{*}{$\sigma^{comp}_y$} & \multirow{2}{*}{Density} &  Rayleigh    \\
                      &                           &                                    &                          &  Damping   \\
$(MPa)$               &                           & $(MPa)$                            & (Dynamic)                & (Dynamic)  \\ \hline
20000                 & 0.3                       & 60                                 & 0                        & 0.1        \\ \hline
\end{tabular}
\end{table}

\begin{figure}
	\centering
	\begin{subfigure}{0.49\textwidth}
	    \centering
        \includegraphics[height=1.45 in]{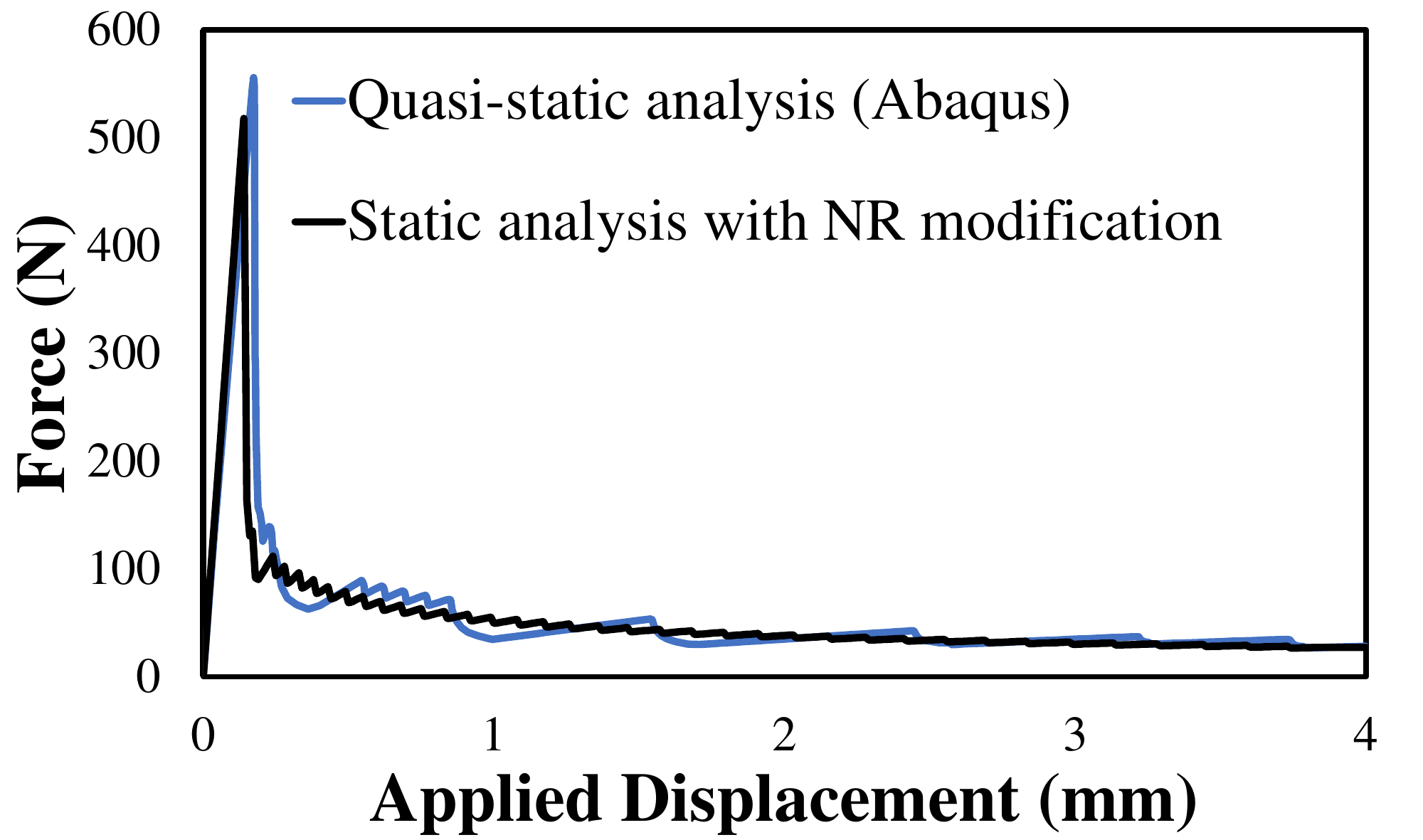}
       \caption{} \label{FIG:22a}
    \end{subfigure}
	\begin{subfigure}{0.49\textwidth}
	    \centering
        \includegraphics[height=1.45 in]{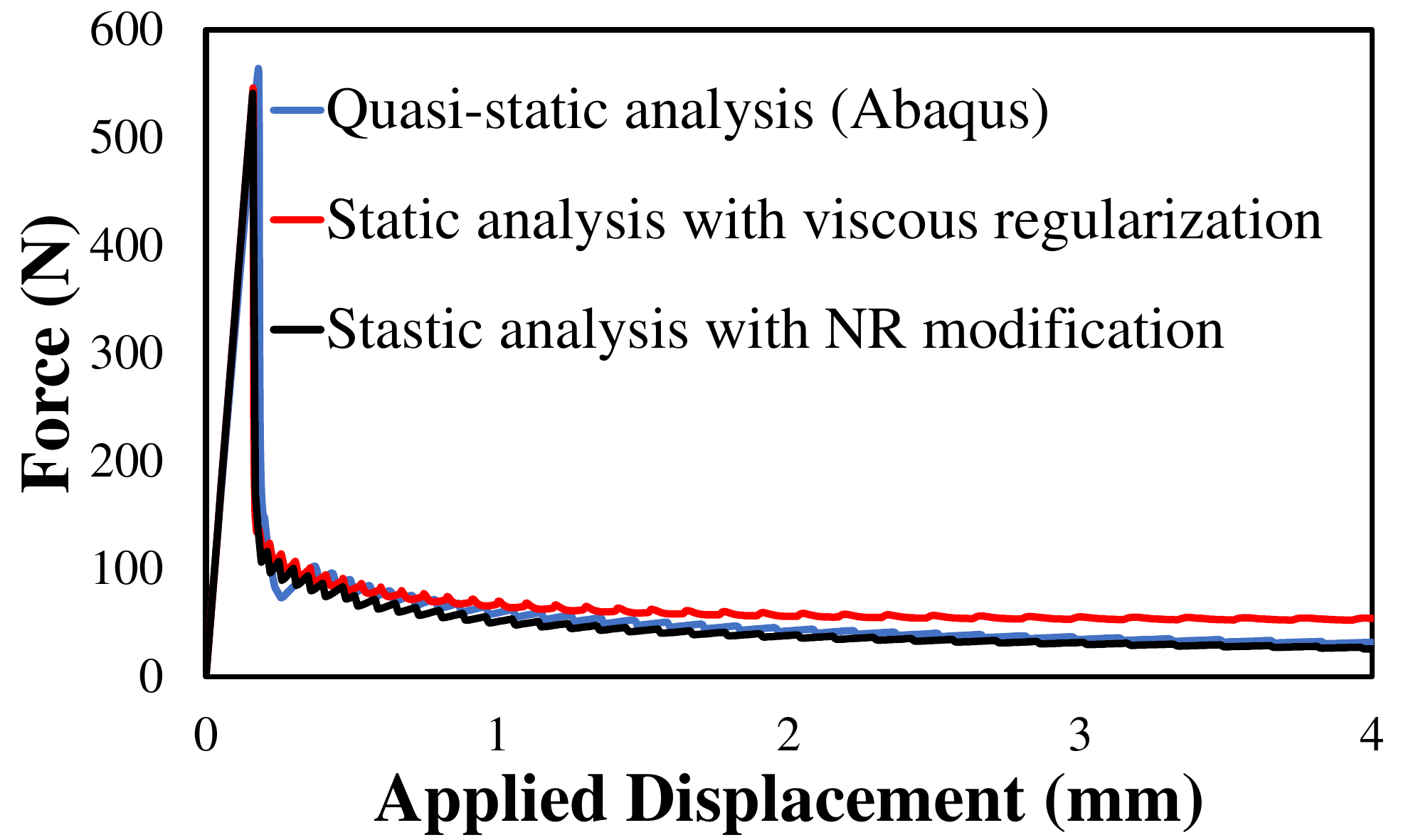}
      \caption{}  \label{FIG:22b}
    \end{subfigure}    
	\caption{Force-displacement responses of the double-cantilever beam model obtained by static and dynamic analyses for the cases of (a) $\nu=0.3$, and (b) $\nu=0$.}
	\label{FIG:double cantilever response}
\end{figure}

\textcolor{blue}{Fig.~\ref{FIG:double cantilever model}~(c)} presents the deformed shape of the double-cantilever beam. \textcolor{blue}{Fig.~\ref{FIG:double cantilever response}~(a)} demonstrates the force-displacement responses for the dynamic analysis and the static analysis with the NR modification. As it can be seen in this figure, the results are slightly different, because Abaqus does not use the Gauss integration method for first-order elements (i.e., isoparametric 4-node plane stress/strain elements). Abaqus uses a method based on single-point integration, which improves the elements' performance and makes them pass the patch test \textcolor{blue}{\cite{abaqus}}. One way to minimize the effect of the different integration methods is to set the Poisson's ratio equal to zero. \textcolor{blue}{Fig.~\ref{FIG:double cantilever response}~(b)} illustrates the force-displacement responses for the case where the Poisson's ratio is zero. As it can be seen, the results from the static analysis with NR modification matches very well with that of the dynamic analysis, which demonstrates the validity and accuracy of the proposed method. No time step refinement was required throughout the analysis when the proposed method was used. As seen in \textcolor{blue}{Fig.~\ref{FIG:double cantilever response}~(b)}, the analysis with the viscous regularization method is slightly inaccurate compared to the dynamic analysis, although 113 time step refinements were required. 

\begin{figure}
	\centering
		\includegraphics[height=1.45 in]{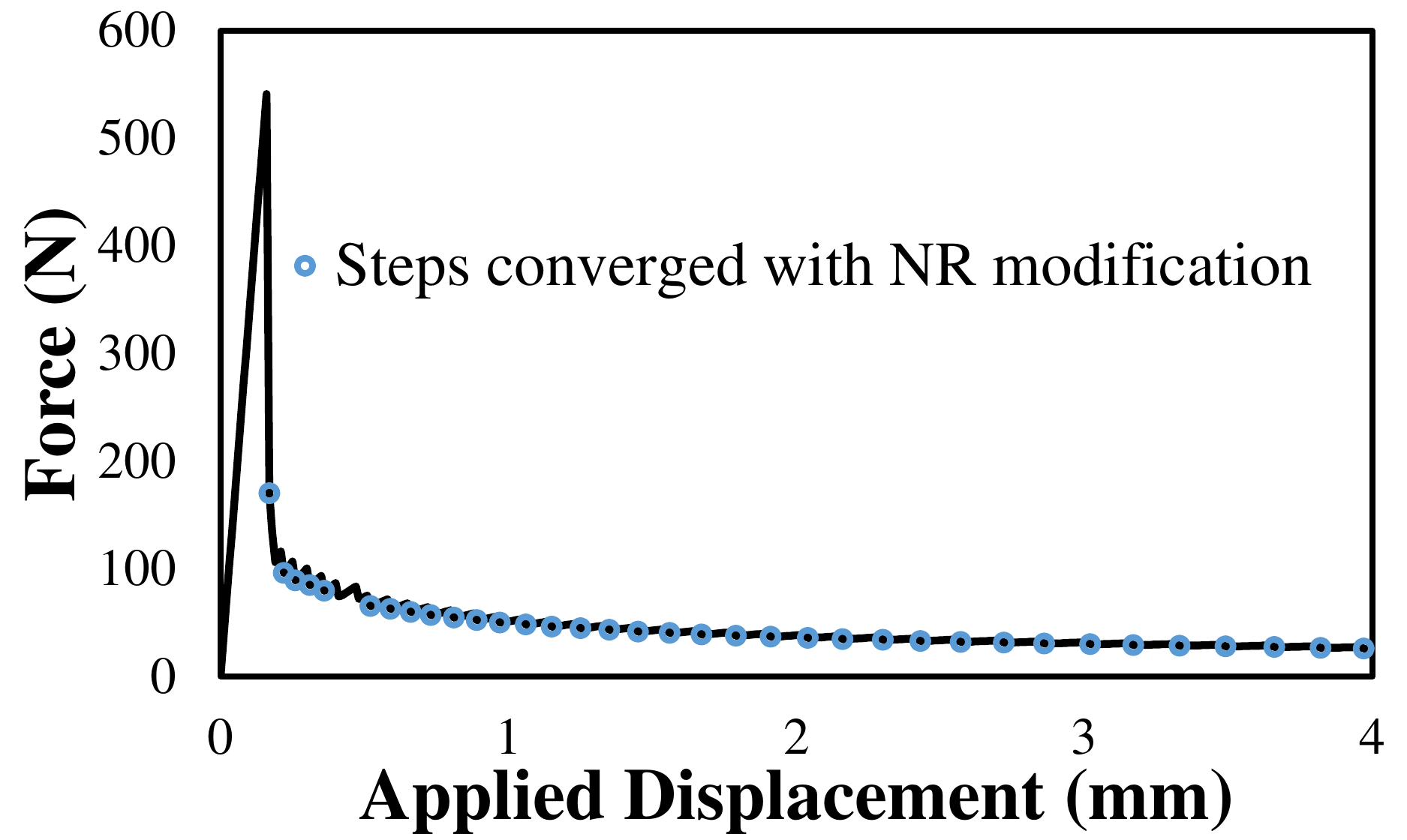}
	\caption{Force-displacement response of the double-cantilever beam model with $\nu=0$, where the blue circles demonstrate the steps converged using the modification method.}
	\label{FIG:double cantilever corrective steps}
\end{figure}

\textcolor{blue}{Fig.~\ref{FIG:double cantilever corrective steps}} illustrates the steps with convergence difficulty that only could converge with the application of the proposed NR modification method for the case of $\nu=0$. A total of 35 steps had convergence difficulty. An average of 6 iterations were taken towards convergence each time the modification method was applied. 
The results presented in \textcolor{blue}{Fig.~\ref{FIG:double cantilever response}} and  \textcolor{blue}{Fig.~\ref{FIG:double cantilever corrective steps}} confirms that the proposed method is both accurate and computationally efficient.


\subsection{2-D bar with a circular interface}
\begin{figure*}
    \centering
	\begin{subfigure}{0.3\textwidth}
	    \centering
        \includegraphics[width=\textwidth]{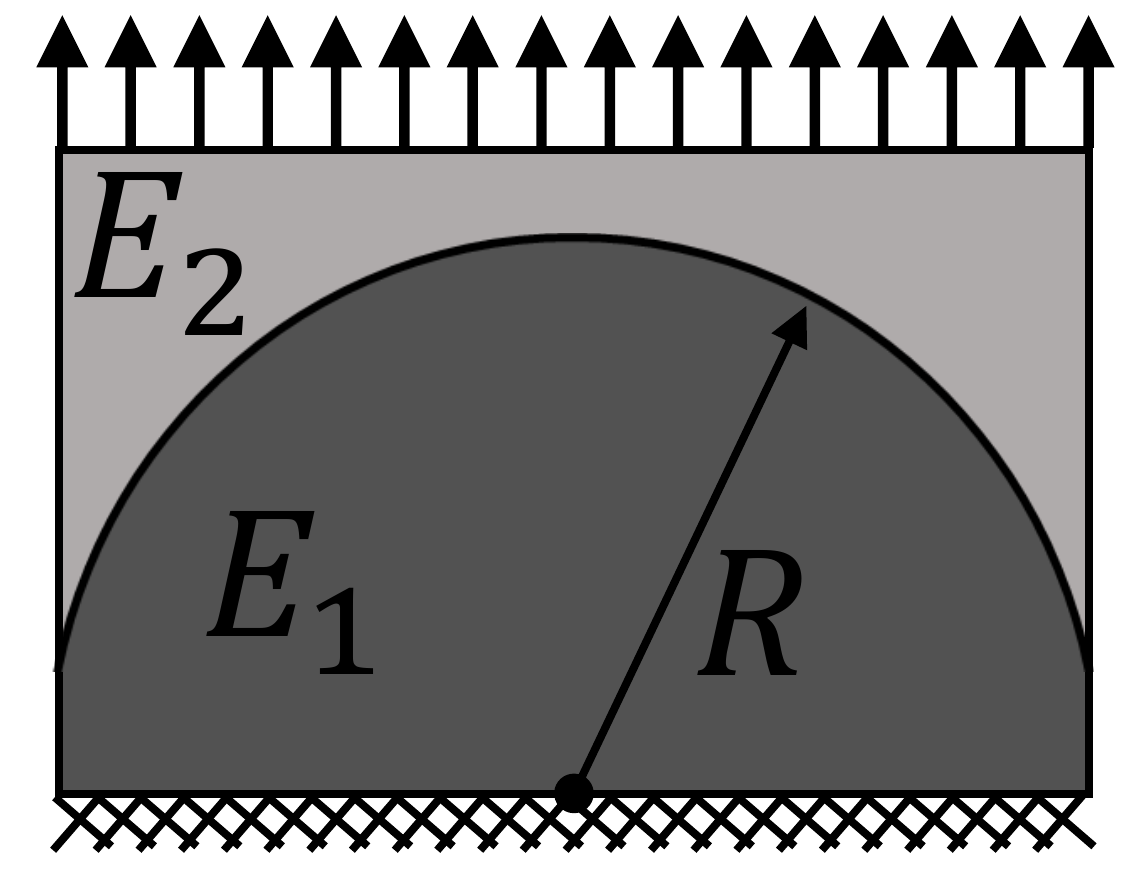}
        \caption{} \label{fig:14a}
    \end{subfigure}
	\begin{subfigure}{0.6\textwidth}
	    \centering
        \includegraphics[width=\textwidth]{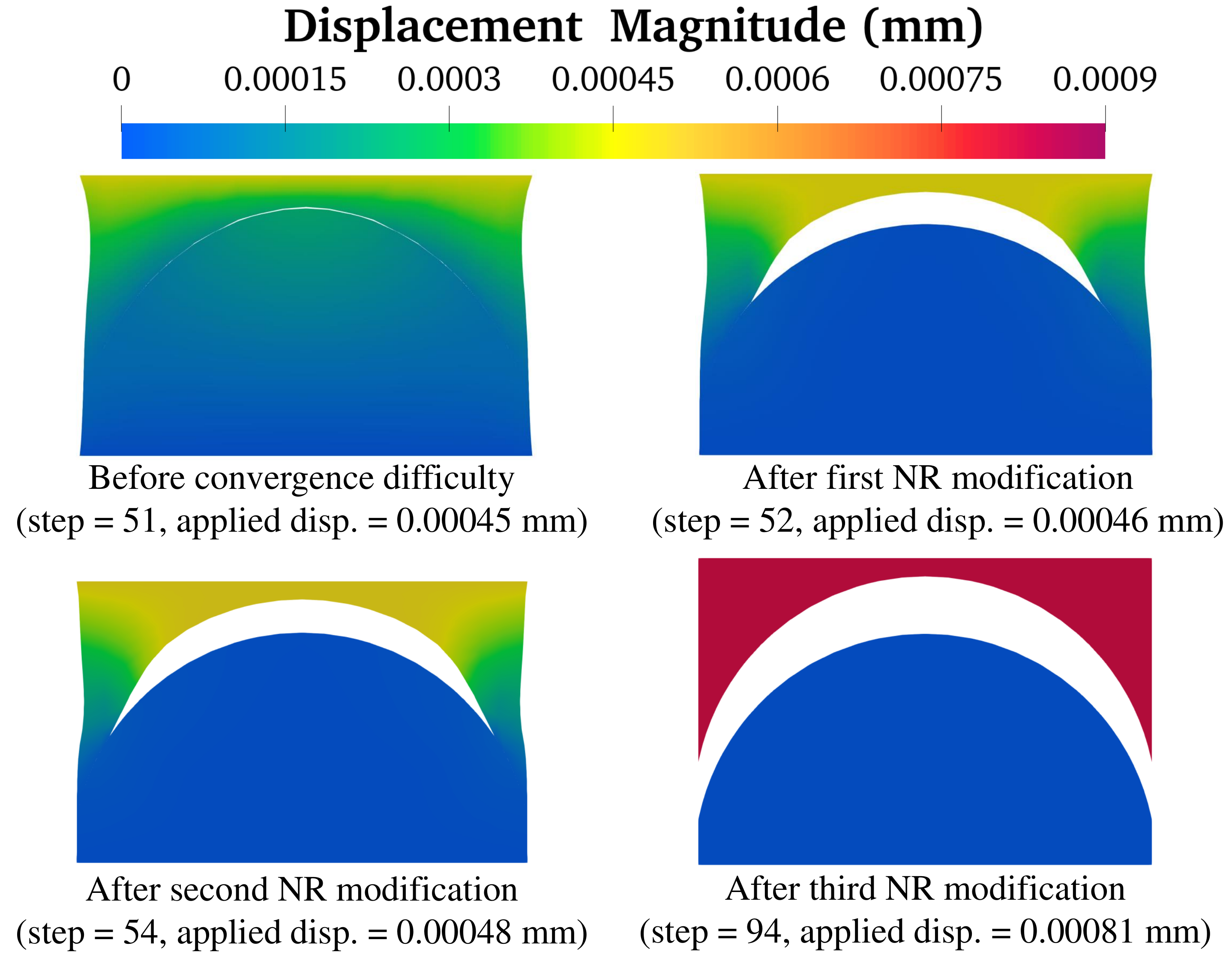} 
        \caption{} \label{fig:14b}
    \end{subfigure}
	\\
	\begin{subfigure}{0.51\textwidth}
	    \centering
        \includegraphics[width=\textwidth]{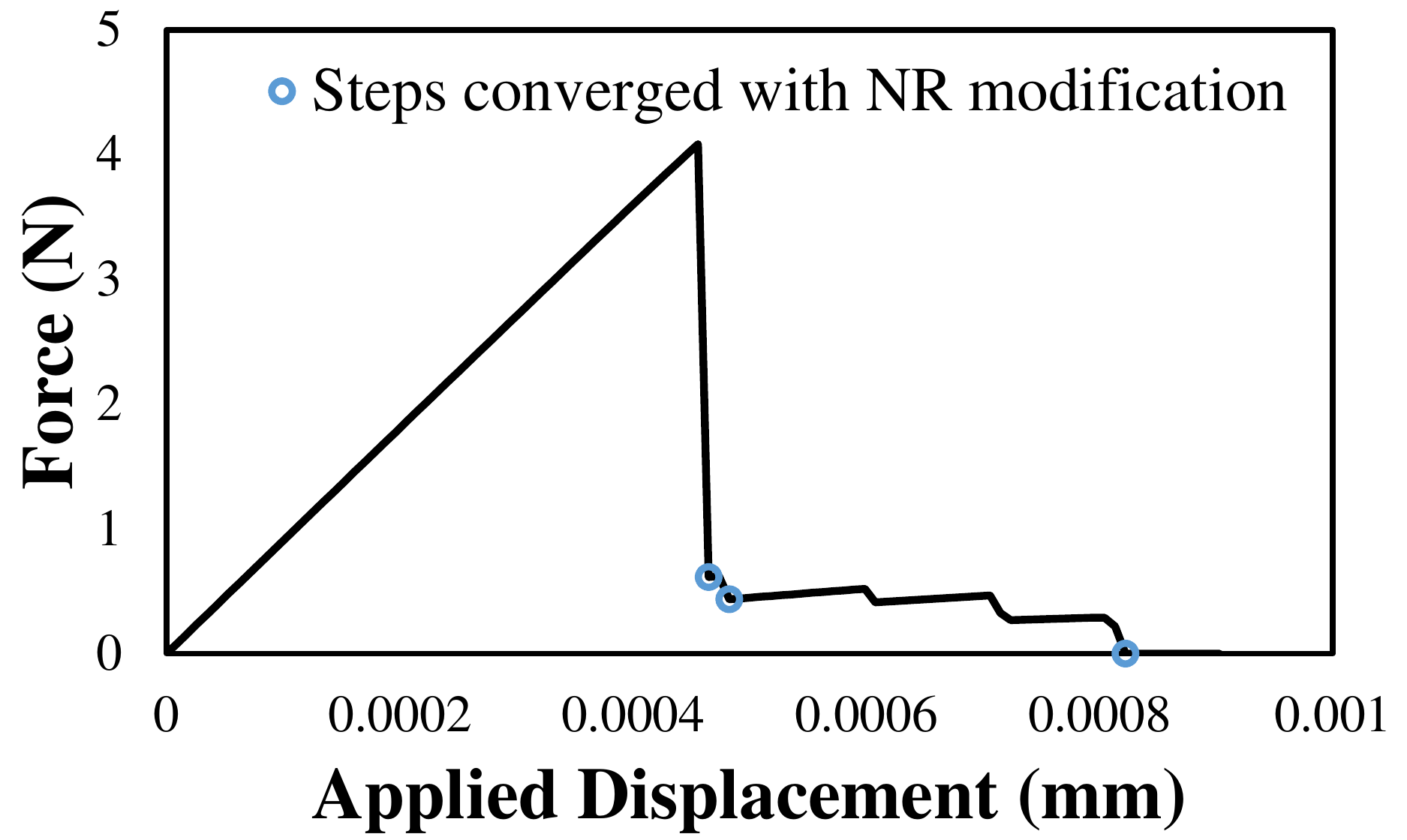}
        \caption{} \label{fig:14c}
    \end{subfigure}
	\begin{subfigure}{0.48\textwidth}
	    \centering
        \includegraphics[width=\textwidth]{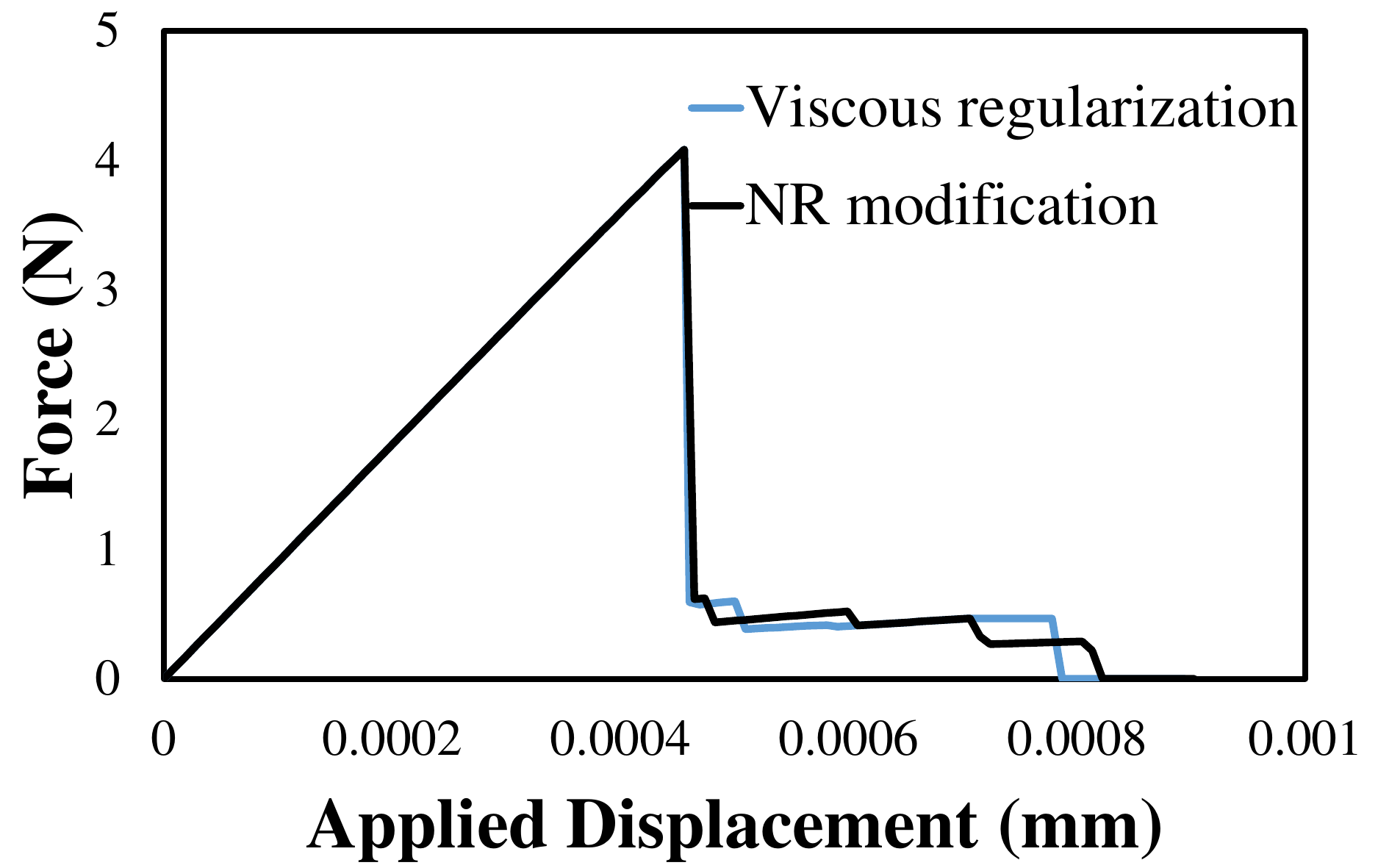}
        \caption{} \label{fig:14d}
    \end{subfigure}

	\caption{The 2-D bar with a circular interface example: (a) boundary conditions, (b) deformed shapes at different loading stages when NR modification is used, (c) displacement steps converged with the NR modification method, and (d)  force-displacement responses for the cases of the NR modification and viscous regularization methods.}
	\label{FIG:curved interface}
\end{figure*}

The third example is a 2-D bar composed of two parts with different material properties connected via a circular interface, as illustrated in \textcolor{blue}{Fig.~\ref{FIG:curved interface}~(a)}. The geometry and boundary conditions were designed to force mixed-mode debonding. A displacement of $0.001\;mm$ was applied to the top boundary in 100 increments while the bottom boundary was fixed. The exponential function presented by \textcolor{blue}{Eqn.~\ref{eq:1}} was used as the CZM's constitutive behavior. \textcolor{blue}{Table \ref{curved interface example properties}} presents the properties of the designed model. It must be noted that Abaqus uses a different normal/tangential coupling scenario for the interface than the one used in the developed FE framework in this work (i.e., \textcolor{blue}{Eqn.~\ref{eq:2}}). Hence, the result from the static analysis cannot be compared with that of a dynamic analysis using Abaqus. The results are just compared to another static analysis which uses the viscous regularization method to address the convergence issue.

Throughout the analysis, 3 steps had convergence difficulty. \textcolor{blue}{Fig.~\ref{FIG:curved interface}~(b)} illustrates the deformed shapes of the model corresponding to those problematic steps. As seen in the figure, the interface debonding prior to the first convergence difficulty (i.e., step 51) was small and no interface element failure occurred. The next displacement increment, which required the application of the NR modification method for the convergence, resulted in an instantaneous failure of several interface elements. The last step with convergence difficulty led to the complete separation of the top and bottom parts in the model. \textcolor{blue}{Fig.~\ref{FIG:curved interface}~(c)} depicts the force-displacement response when the NR modification method was used. The problematic steps which could converge with the application of the proposed method are also marked on the response plot. An average of 5 iterations was taken towards convergence each time the NR modification was applied.

\textcolor{blue}{Fig.~\ref{FIG:curved interface}~(d)} depicts the force-displacement responses using the proposed NR modification versus the viscous regularization method. The comparison indicates that the results from both studied methods very well match up. The decrease in force after the first NR modification is the same for both methods. However, the post-failure responses are slightly different, which is deemed to be a result of the viscous regularization's negative effect on the accuracy. No time step refinement was required for the convergence of the problematic steps when the proposed method was used. The viscous regularization method, however, required six time steps refinement. This example once again demonstrates that the proposed method is more computationally efficient compared to the viscous regularization method.

\begin{table}
    \centering
    \caption{Dimensions and material properties for the 2-D bar with a circular interface.} \label{curved interface example properties}
\begin{tabular}{|c|c|c|c|c|}
\hline
\multicolumn{5}{|c|}{\multirow{2}{*}{Material Properties}}                                  \\
\multicolumn{5}{|c|}{}                                                                      \\ \hline
$E_1$          & $E_2$          & $\nu_1$        & $\nu_2$         & $\sigma_y^{comp}$      \\
$(MPa)$        & $(MPa)$        &                &                 & $(MPa)$                \\ \hline
10000          & 1000           & 0.3            & 0.3             & 60                     \\ \hline
\multicolumn{3}{|c|}{\multirow{2}{*}{Dimensions}} & \multicolumn{2}{c|}{Cohesive Properties} \\
\multicolumn{3}{|c|}{}                           & \multicolumn{2}{c|}{(exponential model)} \\ \hline
Height         & Width          & R              & $\sigma_c$      & $\delta_c$             \\
$(mm)$         & $(mm)$         & $(mm)$         & $(MPa)$         & $(mm)$                 \\ \hline
0.0065         & 0.0036         & 0.0033         & 1000            & 0.00005                \\ \hline
\end{tabular}
\end{table}

\section{Conclusion}
This paper explained the root of the convergence difficulty in CZMs. It was shown that the convergence issue occurs due to numerical instability, which results from the NR iterations entering an infinite cycle. The instability arises in the regions where the material's stiffness is smaller than the CZM's local tangent stiffness. Subsequently, the performance of the two frequently used methods for addressing the convergence difficulty (i.e., viscous regularization method and decrease of global displacement method) was evaluated. It was argued that the methods are either inaccurate, complex, or computationally expensive. 

A simple novel method for addressing the convergence difficulty was proposed, which involves a modification to the starting point of iterations in the NR method. It was schematically shown that if the opening displacement of the problematic PNIEs is increased by an arbitrary large value, the convergence issue is addressed. It was explained that the proposed method does not depend on the debonding mode and the material behavior if the secant stiffness defines the materials' constitutive behavior. The detection of the problematic PNIEs was achieved through the introduction of an interval for the starting point of iterations corresponding to problematic PNIEs. The process of detecting and modifying the problematic PNIEs was explained in detail. Three application examples were then presented to evaluate the performance of the proposed method compared to those of the viscous regularization method and dynamic-implicit analysis. The dynamic analysis showcased the precision of the proposed method. The proposed method outperformed the viscous regularization method in terms of both accuracy and computational efficiency. The functionality of the technique on more complex geometries that undergo mix-mode debonding was also verified. The proposed method can eliminate the convergence difficulty of CZMs in any application area, while the accuracy and computational efficiency are guaranteed.

\section*{Acknowledgments}
The authors like to acknowledge the support for this paper by the Virginia Tech Start-up funding and National Science Foundation award (Grant No. 1853893). 

\bibliography{Paper}
\end{document}